**ESCAPING THE STATE OF NATURE: A HOBBESIAN APPROACH TO COOPERATION IN MULTI-AGENT REINFORCEMENT LEARNING**

A Thesis Presented

By

William F. Long

To

The Departments of Computer Science and Government

in partial fulfillment of the requirements

for a degree with honors of Bachelor of Arts

Harvard College

March 2019

**Acknowledgments**

This thesis is the product of an incredible network of supporters and advisors. I owe a huge debt of gratitude to my advisors Prof. Michael Rosen, for his invaluable feedback and tireless editing, and Prof. David Parkes and Max Kleiman-Weiner, for their patient direction and generous availability. The work here is dedicated to my parents, without whose love and countless sacrifices it would not have been possible. And as with all things, *ad majorem Dei gloriam.*


## Abstract

Cooperation is a phenomenon that has been widely studied across many different disciplines. In the field of computer science, the modularity and robustness of multi-agent systems offer significant practical advantages over individual machines. At the same time, agents using standard independent reinforcement learning algorithms often fail to achieve long-term, cooperative strategies in unstable environments when there are short-term incentives to defect. Political philosophy, on the other hand, studies the evolution of cooperation in humans who face similar incentives to act individualistically, but nevertheless succeed in forming societies. Thomas Hobbes in *Leviathan* provides the classic analysis of the transition from a pre-social State of Nature, where consistent defection results in a constant state of war, to stable political community through the institution of an absolute Sovereign. This thesis argues that Hobbes's natural and moral philosophy are strikingly applicable to artificially intelligent agents and aims to show that his political solutions are experimentally successful in producing cooperation among modified Q-Learning agents. Cooperative play is achieved in a novel Sequential Social Dilemma called the Civilization Game, which models the State of Nature by introducing the Hobbesian mechanisms of (1) opponent learning awareness and (2) majoritarian voting, leading to (3) the establishment of a Sovereign.[1]


---

[1] The code implementation for this research is openly available at www.github.com/wlong0827/state_of_nature

# TABLE OF CONTENTS



I.    <u>Introduction</u>

**Overview**

The central goal of the field of Artificial Intelligence is to produce fully autonomous agents that interact with their environments to learn optimal behaviors, improving over time through trial and error. A principled mathematical framework for generating that experience-driven learning is reinforcement learning (RL) which teaches software agents to take *actions* in an *environment* so as to maximize some notion of cumulative *reward*. Incredible strides have been taken with recent state-of-the-art RL algorithms like deep Q-networks and asynchronous advantage actor-critic that enable intelligent agents to master previously intractable problems[2] including, famously, playing graphical video games[3] and beating humans at Go[4]. But as we apply AI to larger and more complex problems like coordinating entire transportation or manufacturing infrastructures, it becomes increasingly untenable for a single agent to master an entire problem by itself.

Consequently, multi-agent systems (MAS) is one of the most important frontiers of research within the field of AI. Research in this subfield is concerned with how to produce desired global properties in "heterogeneous groups of autonomous agents that pursue partially conflicting goals in an autonomous fashion."[5] Even when each of a group of agents has identical interests – for example, when acting on behalf of a single user or organization – physical or computational considerations can make it desirable to have these agents make decisions independently.[6] With the advent of autonomous vehicles and the explosive growth of software

systems, MAS offer solutions to enable intelligent agents, while pursuing their own objectives (e.g. driving a car from point $A$ to $B$), to interact with other independent agents and humans in a shared environment in a way that leads to optimal global behavior (e.g. minimizing total travel time) without requiring external coordination.

When we place RL agents in a MAS setting, however, we get the problem of cooperation. Traditional learning algorithms are self-regarding and have only the objective of optimizing their behavior so as to maximize their own cumulative reward. So, when placing multiple self-interested agents in a shared environment, often there is divergence between what is best for the individual and what is best for the group. Such scenarios are called *social dilemmas* and they are a matter of interdisciplinary interest from game theory and economics[7] to biology[8] and sociology[9]. Multi-agent reinforcement learning faces an additional challenge because of the unique *perception-action-learning* feedback loop within which each individual agent operates. When the reward received by an agent in a state is a function of not only its action, but the actions of other agents beyond its control, it can have difficulty determining the optimal policy to follow. As a result, a research area of critical importance focuses on how to overcome these obstacles and achieve cooperation in order to make MAS feasible. In the last several decades, this field of research has developed "a vast array of techniques for cooperation and collaboration as well as for agents to handle adversarial or strategic situations," but even so, "current generation agents are unlikely to meet this new challenge except in very simple situations."[10]

The field of political philosophy, likewise, is concerned with the cooperation of autonomous, self-interested individuals in pursuit of common goods. Human society, or political community, is the vehicle by which we achieve these collective goods and, consequently, it is said that "society is the complete political good."[11] While we often take the fact of human cooperation within the nation-state for granted, political philosophers have long argued about whether humans naturally form political communities and which preexists the other. Aristotle summarizes one view when he claims that **"**men journey together with a view to particular advantage, and by way of providing some particular thing needed for the purposes of life, and similarly the political association seems to have come together originally, and to continue in existence, for the sake of the general advantages it brings."[12]

Thomas Hobbes, one of the central modern political philosophers and the father of modern liberalism, on the other hand, takes the opposite view. Gauthier summarizes Hobbes's view as claiming that "individual human beings not only can, but must, be understood apart from society. The fundamental characteristics of men are not products of their social existence."[13] Much of modern political philosophy and social contract theory exists in response to Hobbes' ideas about human nature and interaction. His famous contention is that life in the State of Nature is so terrible that it is overwhelmingly in the best interest of individuals "to confer all their power and strength upon one Man,"[14] the Sovereign, whose exclusive and absolute powers of retributive punishment finally establish stable human society. Once pre-social man comes to realize this necessity, he will desire to subject himself to laws and limitations as long as others are also similarly desirous, and when a majority of them come to this agreement, they ratify the

---

[11] Strauss, L. *What is Political Philosophy? And other Studies*. 1959. 10.
[12] Aristotle, *Ethics* viii. 9.1160a
[13] Gauthier, David, *The Social Contract as Ideology*. 138.
[14] *Id*. at 120.



*Social Contract* and institute the Sovereign. This individualistic train of thinking shaped a strand of later political thought as well, which Macpherson labels "possessive individualism,"[15] in which an individual is conceived as the sole proprietor of his or her skills and owes nothing to society for them.

While it may be unclear at first sight, the problem of cooperation in the fields of MAS and political philosophy overlap in significant ways. Both see a tension between individual and collective interests and both are interested in developing practical mechanisms for moving individuals away from acting on their short-sighted interests, which we might call *defection*, towards a joint policy of *cooperation* that's strictly better for everyone involved in the long-run. It stands to reason, then, that the political solutions that seem to have been successful in promoting civil society for humans might also be conducive to producing cooperation in MAS. Hobbes's theory of the state is well suited to AI because of the applicability of his theory of human nature and the mathematical approach he takes to deducing the formation of the state from those foundational assumptions.

Given these similarities, I suggest that, to the degree that the mechanisms in Hobbes's political philosophy are actually effective in prompting humans to cooperate, those same solutions should be beneficial for teaching MAS to find cooperative policies. The overarching objectives of the thesis, therefore, can be listed as follows:

1) Present an original case for the applicability of Hobbes' philosophy of human nature and moral psychology to AI.

2) Introduce the Civilization Game, a new Markov game modelled on the Hobbesian State of Nature to measure the performance of a MAS composed of RL agents.

---

[15] Macpherson, C. B. *The political theory of possessive individualism: Hobbes to Locke*. 1962.



3) Show that an augmentation of the Civilization Game and standard Q-Learning algorithms with Hobbesian political mechanisms with Opponent Learning Awareness, Majoritarian Voting, and Sovereign reward-shaping produces a significant improvement in the ability of agents to find the long-term cooperative strategy.

This project applies methodologies from these two disparate fields, AI and political philosophy, to synthesize solutions for producing cooperation. This project, beyond its immediate research goals, also seeks to motivate further interdisciplinary exploration that draws on long-standing philosophical ideas about human nature to inform contemporary work on reproducing intelligence in machines. My hope is that the research and solution produced in this thesis might offer a compelling instance of such a synthesis.

## Cooperation and Game Theory

Cooperation is a wide-ranging concept that has been studied across many different disciplines from economics and social science to evolutionary biology and philosophy. Olson famously summarized the central problem by saying that "unless the number of individuals in a group is quite small, or unless there is coercion or some other special device to make individuals act in their common interest, rational, self-interested individuals will not act to achieve their common or group interests."[16] Hobbes likewise agreed that men are not like bees and ants for whom "the Common good differs not from the Private."[17] Even in evolutionary biology, it seems that "Darwinian selection should preclude cooperation from evolving,"[18] because of the constant drive for individual survival.

Popular consensus had generally been that groups of individuals with common interests usually attempt to further those common interests. Groups of individuals with common interests are expected to act on behalf of their common interests much as single individuals are expected to act on behalf of their personal interests. This opinion about group behavior is frequently found in both popular discussions and scholarly work. Such a view has, for example, been important in many theories of labor unions, in Marxian theories of class action[19], and in concepts of "counter-vailing power."[20] It also underpins the Madisonian concept of faction built into the American political system wherein citizens are supposed to be "united and actuated by some common impulse of passion, or of interest, adverse to the rights of other citizens"[21] in a way that translates into policy outcomes.

Olson's thesis motivated both the further study of the dynamics of individual behavior in collective settings and the concept of the organization which "performs a function when there are common or group interests," although organizations often also serve purely personal, individual interests, "their characteristic and primary function is to advance the common interests of groups of individuals."[22] The formation and operation of organizations is often far from optimal, however, as is shown by "the fact that profit maximizing firms in a perfectly competitive industry can act contrary to their interests as a group."[23] In the same way that market pressures cause competing companies to produce and sell goods at lower and lower profit margins though it is against the interest of each company individually, individuals with aligned common interests may fail to realize them in an organization because of competing private interests.

So, then, "if the members of a large group rationally seek to maximize their personal welfare, they will *not* act to advance their common or group objectives unless there is coercion to force them to do so"[24] Coercion is ultimately what enables communities to prevent the Tragedy of the Commons (e.g. coordinating communal resource use, according to Ostrom[25]), governments to pursue public goods (e.g. taxing citizens to build highways), and, crucially for Hobbes's political system, enables individuals in the State of Nature to enforce contracts with one another, for "Covenants, without the Sword, are but words, and of no strength to secure a man at all."[26] Indeed, we will see that for Hobbes, once a central coercive power has been established, the situation individuals face in the State of Nature changes dramatically, and their actions become calculated based not on what they think conduces to their self-interest, but what the Sovereign knows to be best for the society.

Game theory offers a powerful, formal setting within which to study the individual vs collective tensions that exist within organizations like human society or MAS. Hobbes is often said to be the father of game theory since his broader project was to resolve the complex whole of political society into simple social interactions. In *De Cive*, he writes:

> For as in a watch, or some such small engine, the matter, figure, and motion of the wheels cannot well be known, except it be taken asunder and viewed in parts; so to make a more curious search into the rights of states and duties of subjects, it is necessary, I say, not to take them asunder, but yet that they be so considered as if they were dissolved; that is, that we rightly understand what the quality of human nature is, in what matter it is, in what not, fit to

> make up a civil government, and how men must be agreed amongst themselves that intend to
>
> grow up into a well-grounded state.[27]

Hobbes attempts to reduce the whole of human existence in the State of Nature into individual, game-theoretic encounters between one individual and another, and by examining these encounters, culminate with the Social Contract, which represents the establishment of civil government. Depending on what kind of dilemma is faced by pre-social humans, we should see a proper objective, function, and scope of the state.

Skyrms argues that "if one simple game is to be chosen as an exemplar of the central problem of the social contract… the most appropriate choice is not the prisoner's dilemma, but rather the stag hunt."[28] The story is first told by Rousseau in 1750: "If it was a matter of hunting a deer, everyone well realized that he must remain faithful to his post; but if a hare happened to pass within reach of one of them, we cannot doubt that he would have gone off in pursuit of it without scruple."[29]

The game is formalized into a game-theoretic environment when (1) the hunters each have just the choice of hunting the hare or the deer; (2) the deer is much more valuable than the hare; (3) the chance of getting the hare is independent of the actions of other hunters and so is called the *risk-dominant strategy;* (4) the chances of getting the deer is directly dependent on the number of cooperating hunters but is the *payoff-dominant strategy*.

---

| Stag Hunt | C | D |
|-----------|-----|-----|
| C | 4, 4 | 0, 3 |
| D | 3, 0 | 1, 1 |

| Prisoners | C | D |
|-----------|-----|-----|
| C | 3, 3 | 0, 4 |
| D | 4, 0 | 1, 1 |

**Figure 1:** Matrix game social dilemmas. A cell of *X, Y* represents a payoff of *X* for the row player and *Y* for the column player. Each player chooses either to cooperate (C) or defect (D). In the Stag Hunt, agents defect out of fear of a non-cooperative partner whereas in the Prisoner's Dilemma, agents motivated by both fear and greed.

Represented in this way, the stag hunt interaction can be seen in many other places. David Hume postulated a similar dilemma in his *Treatise on Human Nature*: "Two neighbors may agree to drain a meadow, which they possess in common; because 'tis easy for them to know each other's mind, and each may perceive that the immediate consequence of failing in his part is the abandoning of the whole project. But 'tis difficult, and indeed impossible, that a thousand persons should agree in any such action."[30] Hume observed that cooperation in the stag hunt is consistent with rationality, but that the viability of cooperation depends on mutual beliefs and trust. Consequently, the Stag Hunt is also referred to equivalently as the Assurance Game. Rational agents are pulled in one direction by considerations of mutual benefit and in the other by considerations of personal risk.[31] For these reasons, the greater the number of players or agents in the game, the more difficult it is to achieve cooperation.

In this way, the problem of cooperation in the State of Nature can be seen as a Stag Hunt dilemma between individuals where working together will result in common benefit but cooperating when others refuse may lead to personal risk. In the Hobbesian account, then, both the State of Nature and the state of society can be called *Nash equilibria* since it's always better to cooperate when others cooperate and to defect when others defect. Thus, formulated in game-theoretic terms, the fundamental question of both Hobbesian political philosophy and MAS cooperation becomes: "how do we get from the hunt-hare equilibrium to the hunt-stag equilibrium"?

---

[30] Hume, David, *A Treatise of Human Nature*, Book III, Pt II, Section VII, 538.
[31] *Supra*, Skyrms at 3.

## Multi-agent Systems

A *multi-agent system* is defined as the structure in which multiple agents share a common environment, each interacting with every other and having its internal state changed by its own actions as well as those of others. An agent can be a physical or virtual entity that can act, perceive its environment and communicate with others, is autonomous and has skills to achieve its goals and tendencies; even human can be considered components of a MAS[32]. A MAS contains an environment, objects and agents, relations between all the entities, a set of operations that can be performed by the entities and the changes of the universe in time due to these actions.[33] Because each agent has no control over the actions of others and "no global control is applied to the participating agents, it has a subjective view of the evolution of the world."[34] Since this "evolution" is partly a function of actions beyond any individual agent's control, we can understand why a learning algorithm might find it difficult to converge to long-term cooperative behavior since an agent's action only imperfectly maps to a resulting outcome or reward, meaning that it must learn to reason about other agents' actions and intentions.

---

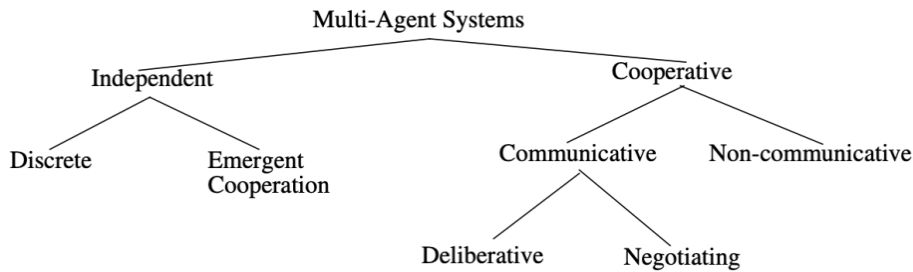

**Figure 2:** A typology of cooperation in MAS. The type that is aimed at in this experimental setting is emergent cooperation since each agent in the Civilization Game maintains its own independent agenda, self-preservation. (Doran et al.)

Cooperation is "one of the key concepts which differentiates MAS from other related disciplines such as distributed computing, object-oriented systems, and expert systems."[35] A MAS is independent if each agent pursues its own agenda[36] independently of the others. It is discrete if it is independent and the agendas of the agents bear no relation to one another; these systems involve no cooperation. On the other hand, a system's behavior is considered emergent if it can only be specified using descriptive categories which are not to be used to describe the behavior of the constituent components.[37] This behavior maintains an intrinsically "subjective aspect by directly depending on the cognitive properties of an observer."[38] For example, the puck gathering robots of Beckers et al.[39] represent an independent system, each following the agenda of moving in a straight line till an obstacle is encountered and then changing direction. Puck gathering is an emergent behavior of the system in that, from an observer's viewpoint, the agents appear to be working together, but from the agent's viewpoint they are not. They are simply carrying out their own individual behavior. This emergent cooperation is ultimately the objective

being pursued for the MAS playing the Civilization Game, and the relevant benchmark for success will be something measurable only at the system level.

When MAS find successful strategies for cooperation, "the interactions of individual agents possess a value-add compared with each simple agent capability."[40] Specifically, these advantages include natural implementations of problems that ask for distributed data and control, greater robustness, scalability, and reusability. Looking ahead to an increasingly automated future, MAS will also offer a valuable paradigm for human-computer cooperation.

## Sequential Social Dilemmas

Before delving into the agents themselves, we'll lay out the structure of the environment within which they will operate. In order to meaningfully explore the concept of cooperation in MAS, agents must face a social dilemma: an environment or scenario which exposes the tension between collective and individual rationality. If individual objectives align with the common good, then cooperation is achieved trivially. Historically, the theory of repeated general-sum matrix games has provided a powerful framework for understanding cooperation through a mathematical model of social dilemmas and has produced famous strategies like Tit-for-Tat in Iterated Prisoner's Dilemma.[41]

---

|   | C | D |
|---|---|---|
| C | R, R | S, T |
| D | T, S | P, P |

**Figure 3:** Outcome variables *R*, *P*, *S*, and *T* are mapped to the cells of a generic game matrix. Incentives are calculated as *Fear* = *P* − *S* and *Greed* = *T* − *R* and games are classified based on those incentives that exist in a game.

At each stage of the game, there are precisely four possible outcomes: the player will receive a reward for mutual cooperation, *R*, a punishment arising from mutual defection, *P*, a "sucker" outcome obtained by the player who cooperates with a defecting partner, *S*, and a temptation reward achieved by defecting against a cooperating player, *T*. A matrix game is a social dilemma when its four payoffs satisfy the following social dilemma inequalities[42]:

1) $R > P$. Mutual cooperation is preferred to mutual defection

2) $R > S$. Mutual cooperation is preferred to being exploited by a defector

3) $2R > T + S$. Mutual cooperation is preferred to an equal probability of unilateral cooperation and defection

4) Either of greed: $T > R$ or fear: $P > S$. Mutual defection is preferred over being exploited

These Matrix Game Social Dilemmas (MGSD) have been fruitfully utilized to study a wide variety of phenomena and scenarios in social science and biology[43], but game-theoretic frameworks like the two-player Iterated Prisoner's Dilemma, Stag Hunt or other multi-player matrix games[44] necessarily treat the choice to cooperate or defect as a discrete, atomic action. Either a hunter will choose to hunt hare or to hunt stag. In real-world social dilemmas, however, these choices are temporally extended. Cooperativeness is a property that applies to policies not elementary actions and consequently exists on a graded spectrum. Decisions to cooperate or defect are made with only impartial information about other players' activities and the state of the world.[45] A real hunter scouts, tracks, runs, and shoots, but we don't label these actions as intrinsically cooperative or defecting. Rather when we observe the broader policy of observing other hunters pursuing the stag and then joining in the chase, we call the hunter's strategy cooperative, and when he chooses the easier route to the hare, disregarding the actions and intentions of his fellow hunters, we call it defecting.

---

[42] Macy, Michael and Flache, Andreas. *Learning dynamics in social dilemmas*. 2002. 7229-7236.
[43] *Supra*, Ale et al.
[44] Broom, M, Cannings, C, and Vickers G. T. *Multi-player matrix games*. 1997. 931-952.

Recent research in social dilemmas centers around Markov game-based Sequential Social Dilemmas (SSD) that treat cooperativeness as "a property that applies to policies, not elementary actions" and requires "agents to learn policies that implement their strategic intentions."[46] Because cooperation and defection are measured through observed behavior and not simply through binary choice, they offer the most powerful framework today for measuring true cooperation in realistic ways.[47] SSDs are a recent innovation on Markov Games that share the mixed incentive structure of MGSDs but also require agents to learn policies that implement their strategic intentions.

A Markov game is formally defined, in the two-player case for simplicity, as $M = (S, O, A_1, A_2, T, r)$ where $S$ is the set of states and $O$ is the observation function $O: S \times \{1, 2\} \to \mathbb{R}^d$ specifying each player's partial $d$-dimensional view of the current state of the world, $A_1$ and $A_2$ denoting the two players' sets of legal actions, a transition function $T: S \times A_1 \times A_2 \to \Delta(S)$ where $\Delta(S)$ is the set of discrete probability distributions over the state

space $S$, and the reward function $r_i : S \times A_1 \times A_2 \rightarrow \mathbb{R}$ for player $i$. If we further let each player's observation function be defined as $O_i = \{o_i \mid s \in S, \ o_i = O(s, i)\}$, then to choose actions, each player uses a policy $\pi_i : O_i \rightarrow \Delta(A_i)$. Formalized in this way, it becomes apparent that matrix games are simply the special case where the state is perfectly observable ($O_i(s) = s$) and the size of the state $|S| = 1$. In general, we can split the universe of all possible legal policies into two disjoint sets of cooperative and defecting policies $\Pi^C$ and $\Pi^D$ where there is no guarantee that $\Pi^C \cup \Pi^D = \Pi$ since for sequential behavior, cooperativeness will usually be a graded property. Thus, we define these two sets by thresholding a continuous social behavior metric $\alpha : \Pi \rightarrow \mathbb{R}$ with values $\alpha_c$ and $\alpha_d$ such that $\alpha(\pi) < \alpha_c \leftrightarrow \pi \in \Pi^C$ and $\alpha(\pi) > \alpha_d \leftrightarrow \pi \in \Pi^D$. This behavior metric will represent our key benchmark for achieving cooperation.

Formally, we can define an SSD as the tuple $(M, \Pi^C, \Pi^D)$ when the Markov Game $M$ satisfies the condition that there exist states $s \in S$ for which the induced empirical payoff matrix satisfies the social dilemma inequalities above. SSD games studied include the Gathering game, where players choose between shooting other players which temporarily disables them and gathering apples located around a grid-like map, and the Wolfpack game, where players can choose to capture prey alone or capture together with other players to obtain an extra bonus.[48] Both these games expose tensions between immediate short-term rewards and optimal cooperative behavior that leads to long-term bonuses but requires adjusting one's policies to coordinate with others. I introduce here a new SSD, the Civilization game, that faithfully models the dynamics that Hobbes posits in the State of Nature and enables the study of a more complex range of incentive structures than previous SSDs.

---

[48] *Supra*, Leibo et al.



# Reinforcement Learning

Turning now to the mechanics of the agents within a MAS, we see that there are three broad approaches to learning: (1) supervised learning where the correct output is provided by a teacher during a training phase; (2) un-supervised learning where there is no knowledge of the correct output but where it is possible to distinguish equivalence classes in the input itself to guide the learning process; and (3) reward-based or reinforcement learning where only a feedback is provided by the environment on the utility of the performance. Of these, the theory of reinforcement learning provides the most natural framework, deeply rooted in psychological[49] and neuroscientific[50] perspectives on animal behavior and how agents can optimize their control of an environment. The family of RL algorithms, as a subset of all machine learning algorithms, is particularly concerned with how to act in a given environment so as to maximize cumulative reward. From this family, the Q-Learning algorithm represents an extremely powerful RL technique because it is model-free, meaning it doesn't require complete knowledge of the environment, and consequently well suited to play partially-observable Markov games.

Q-Learning computes a value for every state and action combination with a function $Q : S \times A \rightarrow \mathbb{R}$ where the magnitude of the Q-value is proportional to the perceived value of taking action $A$ at state $S$. At any state, a QL agent simply looks at the set of all possible actions that can be taken and, under the epsilon greedy strategy, chooses with some probability $\epsilon$ to take the action with the highest Q-value and takes a random legal action with probability $1 - \epsilon$. This strategy enables the agent to both explore the state space by performing actions it wouldn't otherwise take while also exploiting learned experience. It also embodies the classic explore vs

---

exploit trade-off where taking actions that are known from experience to yield rewards may prevent an agent from finding global maximums.

Procedurally, all Q-values are initially assigned a random value before learning begins. Then, at every time $t$, the agent selects an action $a_t$ from state $s_t$ and receives reward $r_t$ and updates the Q-value according to the Bellman update equation:

$$Q(s_t, a_t) \leftarrow (1 - \alpha)Q(s_t, a_t) + \alpha(r_t + \gamma \cdot \max_a Q(s_{t+1}, a))$$

The parameter $\alpha \in [0,1]$ is the learning rate where a value of $\alpha = 0$ means that the agent essentially learns nothing, exploiting only previous experience, and a value of $\alpha = 1$ means that the agent considers only the most recent experience, ignoring prior knowledge. The discount factor $\gamma \in [0,1]$ determines the importance of future rewards where $\gamma = 0$ makes the agent myopic, or short-sighted, by considering only immediate reward, and $\gamma = 1$ will make it strive towards future reward. Taken altogether, the update equation describes a simple value iteration update ran for a fixed length of time that utilizes a weighted average of old values and new information. We will see later how RL and the learning parameters in particular find a parallel in Hobbes' theory of the nature of human perception and decision-making.

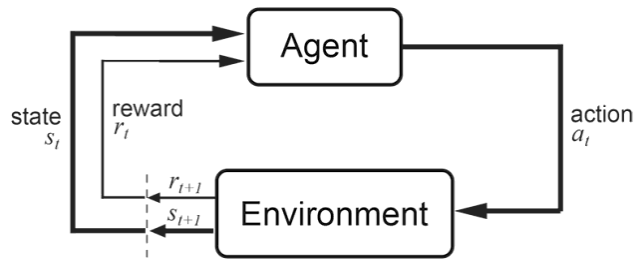

**Figure 4:** The perception-action-learning loop where the action $a_t$ taken translates directly into the reward $r_{t+1}$ and new state $s_{t+1}$. In a MAS, learning becomes unstable because the reward and state are functions of other agents' actions over which a player has no control. (Sutton and Barto)



For Q-Learning in the multi-agent system setting, the $i$-th agent stores its own Q-function $Q_i : S \times A_i \to \mathbb{R}$ in a Q-value table. The policy for that agent can then be written as:

$$\pi_i(s) = \begin{cases} \text{argmax}_{a \in A_i} Q_i(s,a), & p = \epsilon \\ U(A_i), & p = 1 - \epsilon \end{cases}$$

Where $U(A_i)$ denotes a sample of legal actions from the uniform distribution. In independent multi-agent reinforcement learning, each agent regards the actions of others simply as part of the environment. This independence assumption can be considered an assumption of *bounded rationality* in that agents do no reasoning about other agents' learning and it represents a major obstacle for achieving cooperation in MAS. Some promising work has been done recently to make agents aware of opponents' learning[51], and that paradigm will also play an important role in implementing a Hobbesian mechanism for achieving cooperation.

## Hobbes and Leviathan

From what we've considered so far, it might not yet be obvious why the political philosophy of Hobbes would play an intuitive or valuable role in offering new solutions to the problem of cooperation in MAS. Providing some historical and philosophical background on this thinker and his unique ideas will hopefully help make the case for their applicability to the problem at hand.

Thomas Hobbes (1588-1679) was an English political philosopher writing in a tumultuous period of European history. From the French Wars of Religion and the bloody Thirty Years' War to the English Civil War culminating in Cromwell's Protectorate, the 17th century was one characterized by extraordinary instability and savagery. Given this kind of political turmoil, it is not surprising that Hobbes came to hold a view of human beings as creatures who will, if unchecked, inevitably behave violently toward one another. The chief political struggle in England during this period was between the Parliamentarians, who advocated the rights of Parliament, and the Royalists, who supported King Charles I.

---

[51] Foerster, Jakob et al, *Learning with Opponent-Learning Awareness*. 2018

In particular, Charles's claim to the right to imprison nobles *per special mandatum domini reges* (by special command of the King) raised the question of whether the King was above the Law and could therefore subvert it in order to secure the Commonwealth or whether he himself was subject to the Law. Hobbes, seeking to settle this disagreement which was causing so much conflict and bloodshed, sided strongly with the King, presenting his famous argument in Leviathan that the Sovereign embodied the body politic and consequently exercised complete and absolute authority. No entity could behave with a single will and reject its Sovereign; he could not be tried and executed by the authority of 'the people' as Charles I was. Hobbes's royalist friends in 1640-2 were well satisfied with this argument, for it secured to the king all the powers vis-à-vis Parliament which he claimed in those years,[52] although they eventually realized that those same arguments would go on to support obedience to Cromwell's republican government.[53] Hobbes himself was likely a staunch royalist whose radical arguments ended up undermining his own monarchist position.[54]

*Leviathan* is both Hobbes's systematic construction of the social state and consequently, it is the work this thesis will build its cooperative mechanism upon. It is also the culmination of his metaphysical, moral, and natural philosophy begun in *De Cive* and *Elements of Law,* and so these two works will primarily form the basis of the argument motivating the application of

---

[52] *Supra*, *Leviathan* at xix
[53] Skinner, Q. *Conquest and consent: Hobbes and the engagement controversy*. Visions of Politics. 2002. 287-307
[54] Hamilton, J. *Hobbes the royalist, Hobbes the republican*. History of Political Thought. 2009. 411-454.



Hobbesian political philosophy to Artificial Intelligence. To the degree that his natural and moral philosophy is applicable to artificially intelligent agents, the mechanism or process he proposes in his political philosophy for the formation of society should in principle be applicable to producing cooperation in MAS. Consequently, a substantial portion of this thesis will offer an interpretation of his metaphysics and natural and moral philosophy as uniquely fitting for artificially intelligent agents, specifically reinforcement learning-based AI. Finally, on the strength of that comparison, it will implement Hobbes's political solutions as modifications to standard RL to test in the Civilization Game.

Before that, understanding Hobbes's philosophical methodology is important. For Hobbes, philosophy is divided into three branches: natural, moral, and political, each of which study a different kind of "body". His natural philosophy is guided by a fundamental belief in *mechanistic materialism*, claiming that the universe is machine-like, and *strict determinism*, acting according to strict scientific laws of cause and effect and being composed exclusively of material bodies. Similarly for humans, there is no such thing as soul or spirit; rather, everything from our actions, emotions, and states of mind, are series of internal "motions" caused by external sensation. This materialism enables Hobbes to attempt a "science of politics", deriving political truths from fundamental, self-evident principles through a logical, almost mathematical, methodology. He makes a case for political argumentation based on reason, or "ratiocination", alone, since "the principles of politics consist in the knowledge of the motions of the mind."[55]

As Hampton puts it, "Hobbes characterizes his political argument as a 'demonstration' in the style of geometry, in which his political conclusions are derived from his psychological theories, which in turn can be derived from his physiological theories of human behavior."[56] In his biography of Hobbes, John Aubrey explains the encounter that inspired this radical approach:

> He was... 40 years old before he looked upon geometry; which happened accidentally. Being in a gentleman's library..., Euclid's Elements lay open, and 'twas the 47 El. libri I. He read the proposition. 'By G—,' said he (he would now and then swear, by way of emphasis), 'this is impossible!' So he reads the demonstration of it, which referred him back to such a proposition, which proposition he read. That referred him back to another, which he also read. Et sic deinceps, that at last was demonstrably convinced of the truth. This made him in love with geometry.[57]

---

[55] *Supra*, *De Corpore* at 6.74.

Impressed with the incontrovertible conclusions which geometry enabled mathematicians to show, Hobbes set his mind to deriving a system of government in a similar fashion. In his mind, the serious political problems of his time resulted from disagreements over who had the ultimate authority in political and religious affairs.

Consequently, a purely rational methodology could avert the constant quarrels amongst philosophers about the proper structure and scope of government. This system of careful, deductive, one might say, mathematical reasoning, seeks to establish definitively the proper organization and scope of political society. Beginning with Hobbes's fundamental premises about human nature and the universe, he forms a chain of consequences that passes through morality and society before reaching politics. If we can show that Hobbes's premises in natural and moral philosophy hold, then his political prescription for the problem of Cooperation should also prove effective.

It's worth noting that this rationalist epistemology and materialist metaphysics radically contrasted with the Christian Scholastic philosophy of Hobbes's immediate predecessors and to other traditional thinkers like Augustine. Augustine, for example, also defined *civitas* as "nothing

---

else than a multitude of men bound together by some associating tie,"[58] but he emphasized that this was insufficient to distinguish *civitas* from other forms of communal life. It is only when we give due consideration to the religious and moral implications of the adjective "earthly," that the distinctive traits of the *civitas terrena* become plain.[59]

And, for a modern example, Tocqueville insisted that he would "never admit that men form a society by the sole fact that they acknowledge the same leader and obey the same laws; there is a society only when men consider a great number of objects in the same way; when they have the same opinions on a great number of subjects; when, finally, the same facts give rise among them to the same impressions and the same thoughts."[60] Disregarding mores, religion, and common culture as superfluous to the formation of political society, Hobbes was freed to pursue his geometric proof for the foundation of the state, but arguably lost the characteristics that made his subjects uniquely human. But, regardless of whether it represents a compelling model for human nature, I will show that his materialist and rationalist framework can apply to a class of standard AI methodologies.

---

[58] Augustine, *City of God.* XV. 73.
[59] Loetscher, Frederick. *St. Augustine's Conception of the State*. 1935. 16-42.
[60] Tocqueville, Alexis, *Democracy in America.* 598



II.     <u>Methodology</u>

## AI as Hobbesian Man

The field of Computer Science and specifically its subfield Artificial Intelligence offers a unique opportunity for Hobbesian scholarship, because, arguably, the tenets of Hobbes's natural and moral philosophy apply remarkably well to computational intelligent agents. Six core propositions, two from his foundational natural philosophy and four from his intermediate human or moral philosophy, will enable us to both examine this applicability and understand the full geometric logic of Hobbes's political argument for the necessity of the Sovereign:

1) Mechanistic Materialism – *The universe consists solely of sensible material and motion*

2) Strict Determinism – *Action in the universe is entirely determined by material causes*

3) The Moral Vacuum – *Humans can/should do whatever is necessary for self-preservation*

4) Rational Self-Interest – *Humans are rational and driven by subjective interest*

5) Radical Individualism – *Humans should be considered as fundamentally individuals.*

6) Equality of Condition – *Humans are essentially equal in mental and physical capacity*

Hobbes's mechanistic materialism and determinism (the first two propositions) lead him to see human nature as characterized by moral subjectivism, rational self-interest, radical individualism, and equality of condition, which ultimately serves to prove "by experience known to all men and denied by none, to wit, that the dispositions of men are naturally such, that except they be restrained by some coercive power, every man will dread and distrust each other."[61] I shall examine each in turn.

---

[61] *Supra*, *De Cive* at 103.



*Mechanistic Materialism*

For Hobbes, all that exists is physical: "for the Universe, being the aggregate of all bodies, there is no real part thereof that is not also Body, nor anything properly a Body, that is not also part of the Universe."[62] The idea of metaphysical realities like soul or spirit are reducible to physical, dispositional, or imaginative phenomena since substances are corporeal by definition. "Substance and body signify the same thing and therefore substance incorporeal are words, which when they are joined together, destroy one another, as if a man should say, an incorporeal body"[63] Even God Himself is not what is commonly called Spirit, and when we refer to him as such, we do not "signify our opinion of his Nature, but our desire to honor him with such names as we conceive most honorable amongst ourselves."[64] In effect, he deflects man's claims about the reality of the supernatural or metaphysical as simply a pious means of expressing our intentions and opinions.

This materialist premise quickly translates into a mechanical interpretation of human action: since life "is but a motion of limbs, the beginning whereof is in some principal part within, why may we not say, that all Automata have an artificial life? For what is the Heart, but a spring, and the Nerves, but so many strings, and the Joints, but so many wheels, giving motion to the whole both, such as was intended by the Artificer?"[65] Man is no more than a mechanical automaton controlled and impelled by an internal force. And what is that internal force? Mere Imagination, simply the impression of some object upon the senses, the "conception remaining, and by little and little decaying from and after the act of sense,"[66] which he calls "the first internal beginning of all Voluntary Motion." By the movement of the Imagination, "these small beginnings of Motion, within the body of Man, before they appear in walking, speaking, striking, and other visible actions, are commonly called Endeavour" and finally, "this endeavor, when it is toward something which causes it, is called Appetite or Desire"[67] or their opposite, Aversion. In this way, all human desire, and consequent action, is reduced to exclusively material causes through a mechanistic operation. It is in this sense that we can call Hobbes's natural philosophy a mechanistic materialist one.

---

[62] *Supra*, *Leviathan* at 269.
[63] *Id* at 270.
[64] *Id* at 271.
[65] *Id* at Introduction, 9.
[66] Hobbes, Thomas. *The Elements of Law, Natural and Politic*. III. 27.

The implications of this view can be difficult to appreciate in our modern context. Hobbes has overturned the traditional teleological view of the universe begun with Aristotle for whom "every craft and every investigation, and likewise every action and decision, seems to aim at some good" and consequently, the good can be meaningfully defined as "that at which everything aims."[68] Instead of seeing natural ends as the impetus of action, Hobbes's mechanical model sees action merely as the result of blind matter and motion. The Platonic and Christian conception of the incorporeal soul which balances reason with appetite and spirit is rejected. Most pertinently for his political philosophy, Hobbes's reduction of the motivation of all human action to the desires elevates the centrality of "power", the ability to satisfy or obtain the objects of Appetite or avoid the objects of Aversion.[69] And thus, his famous assertion in *Leviathan* that "I put for a general inclination of all mankind, a perpetual and restless desire of Power after power, that ceaseth only in Death."[70]

Hobbes's mechanistic materialist view of the universe is the aspect of his thought most obviously and immediately applicable to AI. Strictly defined, "modern computers are simply

---

[67] *Supra*, *Leviathan* at VI. 38.
[68] *Supra*, *Nicomachean Ethics* at 1094a1.
[69] *Supra*, *Leviathan* at X.
[70] *Id* at XI. 70.



Turing machines that operate on an alphabet consisting of the two symbols '0' and '1'."[71] A Turing machine is simply a mathematical model of computation that defines an abstract machine which manipulates symbols on a strip of tape according to a table of rules. As its creator, British mathematician Alan Turing, described it, "the machine is supplied with a 'tape' (the analogue of paper) running through it and divided into sections (called 'squares') each capable of bearing a 'symbol'."[72] It shuttles the tape back and forth, reading data, making marks on its scratch tape, and inspecting the tape for information on what to do next. Turing proved that such a seemingly naïve and rudimentary contraption, in fact, could embody every imaginable computational operation. Nothing significantly changes when we consider reinforcement learning AI. Every intelligent operation for a Q-Learning agent is simply a computation that considers memories of the quantified "pleasure" or "pain" of being in some state and a further operation to decide what action is best fitted to attain maximal utility. So, then, we can easily see the applicability of Hobbesian materialism to AI.

*Strict Determinism*

Mechanistic materialism is also closely related to strict determinism. If every extant body and potential cause of motion is perfectly contained within the physical, sensible universe, then every subsequent effect or observable motion is reducible to an enumerable set of causes, which are themselves the result of observable causes, and so. Consequently, free will is also cast aside "because every act of man's will and every desire and inclination proceedeth from some cause, and that from another cause, in a continual chain, they proceed from Necessity."[73] Likewise, if we consider that at the root of this chain lies "one first mover, that is, a first and eternal cause of all things, which is that which men mean by the name of God,"[74] then we can also understand why Hobbes might consider himself a sort of compatibilist instead when he explains that "the liberty of man in doing what he will, is accompanied with the necessity of doing that which God will, and no more, nor less"[75] Now, when we apply this Hobbesian premise of strict determinism to human beings, we form a portrait of human cognition as a sole function of a set of past experiences: "conception of the future is but a supposition of the same, proceeding from remembrance of what is past… And that anything hath power now to produce another thing hereafter, we cannot conceive, but by remembrance that it hath produced the like heretofore"[76] The epistemological implications of viewing human learning in such a deterministic way seem to preclude the possibility of transcending our mechanical minds and grasping real truth, and indeed Hobbes himself comes out and explicitly states that "experience concludes nothing universally."[77] Rather, truth, is simply a matter of linguistic convention and "consists in speech and not in the things spoken of."[78] Strict Determinism in Hobbesian natural philosophy, far from being a marginal and inconsequential tenet, clearly has profound implications upon the nature of truth and the capacity for humans to attain it.

---

[71] Weizenbaum, Joseph. *Computer Power and Human Reason*. 1976. III. 73.
[72] Turing, A. M. *On Computable Numbers with an Application to the Entscheidungsproblem*. 1936. 2.
[73] *Supra*, *Leviathan* at XXI. 146.

Similarly, "machines, when they operate properly, are not merely law abiding; they are embodiments of law."[79] In the same way that materialism implies that every action by a physical bodies is a result of a fixed and observable set of physical motion and cause, so for computers, the output of any Turing Machine is simply the conclusion of a set of fixed computational operations upon an input. "Every effective procedure can be reduced to a series of nothing but

commands (i.e. statements of the form 'do this' and 'do that') interlaced with conditional-branch instructions"[80] which simply perform different commands based on whether some claim is true. For Artificial Intelligence, the details of the computational decision process that determines action has been laid out thoroughly above and it is unnecessary to recreate it here.

Thus, there is a striking similarity between Hobbes account of human perception and cognition and the reinforcement learning model. In the same way that Hobbesian man consumes sense information about the world and then translates it into "endeavor" by means of an internal motion, so it is precisely with AI. And just as Hobbesian man is restricted in his cognition to the operation of his imagination which is no more than shadows and memories of past experience, so reinforcement learning, by definition, draws up its actions solely from the outcomes of previous interactions with its environment in order to find the optimal policy of action that results in the obtainment of reward and the avoidance of punishment.

There is, however, one potential deviation from the deterministic nature of that process: the introduction of stochasticity, or randomness, that enables agents to occasionally explore new states that their decision process would otherwise avoid. The conditional comparison of a random number and the exploration parameter $\varepsilon$ in an epsilon-greedy learning strategy determines whether the agent will decide their action based on what has been historically desirable (exploitation) or without factoring in any past experience (exploration). The question of whether true randomness is deterministic is an exceedingly difficult one for which there is no clear consensus amongst philosophers or physicists.

---

[80] *Id,* at III. 97.



*Moral Subjectivism*

Having examined the nature and role of materialism and determinism, we now turn to its effects on Hobbes's moral philosophy. How are humans living in a Hobbesian universe supposed to behave? What injunctions exist in the State of Nature? What are the prerogatives and limitations in such a state?

The first and most natural consequence is moral subjectivity. In precisely the same way that truth is simply a matter of usage and the combination of words into definitions, so "these words of Good, Evil, and Contemptible are ever used with relation to the person that useth them, there being nothing simply and absolutely so, nor any common Rule of Good and Evil to be taken from the nature of the objects themselves."[81] Having dispensed with Aristotle's teleological model of the universe by assuming a materialist framework, Hobbes destroys any independent conception of the Good by precluding metaphysics and then removing the possibility of deriving it from natural ends.

The Laws of Nature that he proposes in *Leviathan* are simply rules discovered by reason for best maintaining one's self-preservation that are binding *in foro interno*, or within one's own thinking and intentions, but not *in foro externo,* or the actual performance in practice. For example, while all men must strive for peace internally, that doesn't necessarily obligate you to work towards it. "For he that should be modest and tractable, and preform all he promises, in such time and place where no man else should do, should be make himself prey to others and procure his own certain ruin, contrary to the ground of all laws of nature, which tend to nature's preservation."[82]

---

[81] *Supra*, *Leviathan* at VI. 39.
[82] *Id* at XV. 107.



This is quite different from what we might normally associate with conscience or natural law, which is both objective, existing independently of one's own conceptions, and binding on external activity. What results, then, is a moral vacuum in the State of Nature with no source of authority or judgment to restrain human action. And so, in the "war of every man against every man, this also is consequent: that nothing can be Unjust. The notions of Right and Wrong, Justice and Injustice have there no place."[83] With these premises securely in place, Hobbes secures himself from the charge that his proposed state of war is prohibited by human conscience and the natural law, since what actually exists are only reason, power, and appetite.

Computers, too, have no equivalent to conscience or natural law traditionally understood. This is perhaps one of the strongest reasons for interpreting AI in Hobbesian terms. In many other systems of natural and moral philosophy, from the Platonic and Aristotelian to the Thomistic, writers assert the existence of both body and soul in humans and that there is an imprint of the Law of God "written on their hearts, while their conscience also bears witness, and their conflicting thoughts accuse or even excuse them."[84] Obviously, such a picture of human nature would be entirely unfit to be modelled by a machine (at least in the existing state of Computer Science), but the mechanistic and deterministic Hobbesian system, by throwing away the concept of natural morality, clears the way for the comparison.

The incapacity of AI to navigate questions of morality is a subject of significant scholarly and policy interest.[85] With the advent of autonomous agents driving cars, determining terms of parole, and fighting wars, policymakers must grapple with "the capacity of algorithms or trained systems to reflect human values such as fairness, accountability, and transparency."[86] The

---

[83] *Id* at XIII. 90.
[84] Rom 2:15
[85] Kleiman-Weiner, M. *Learning a commonsense moral theory*. 2017.
[86] Calo, Ryan. *Artificial Intelligence Policy: A Primer and Roadmap*. 2017. 9.



Hobbesian political system offers a framework for achieving cooperation in multi-agent social dilemmas even without reference to something like moral reasoning.

*Rational Self-Interest*

From the moral vacuum of the State of Nature, follows the primacy of rational self-interest in human action. This Hobbes calls "the Right of Nature, which writers commonly call *Jus Naturale*, the Liberty each man hath, to use his own power, as he will himself, for the preservation of his own Nature… and consequently, of doing anything, which in his own Judgment and Reason he shall conceive to be the aptest means thereunto"[87] Because Law and Right are opposed, with the complete absence of law, natural or civil, in the State of Nature comes the monopoly of right in determining human action.

While it may seem that certain ethical norms like prohibitions upon the murdering of the innocent, for example, may be universally held, Hobbes rejects any intrinsic metaphysical consideration separate from Reason and self-interest. In fact, for him, the variance in these norms across culture and time is greater than the similarity; since a "common standard of virtues and vices," Hobbes says, "does not appear except in civil life; this standard cannot, for this reason, be other than the laws of each and every state."[88] Instead, humans always can and do pursue their self-interest, subjectively defined as that which satisfies their Appetites and removes their Aversions, which in the vast majority of cases (with a few notable exceptions) includes self-preservation.

In this sense, Hobbes falls under the category of what contemporary philosophers might term behavioralism. In place of the prescriptive traditional natural law of Aristotle and Aquinas he substitutes new Laws of Nature derived purely from reason upon the sole premise self-interest which counsel among other things the pursuit of peace, the performance of covenants, and gratitude for benefits rendered, all rationally calculated to achieve one's self-interest. By freeing man from all the constraints of traditional ethical norms, natural law, and conscience, Hobbes removes any factors from the State of Nature that would have been outside the scope of his geometric and deductive methodology. Most importantly, the absence of objective Justice to arbitrate human interaction and check each man's self-interest in the State of Nature allows Hobbes to paint it as a state of complete license to invade a neighbor's territory or renege on one's vows, leading ultimately to continual fear and reprisal.

---

[87] *Id* at XIV. 91.
[88] Hobbes, Thomas. *De Homine*. B. Gert, Indianapolis: Hackett Publishing, 1991. 69.

The decision process for AI agents takes as input the current state of the world (e.g. my location relative to other agents, how much territory I occupy, whether I've been invaded by another, etc.) and compares it to the "memory" of past experience taking action in that state. In the standard case where the agent's objective function is to maximize its own reward, it considers only that single factor when choosing which action to take from a given state. Just as for Hobbesian man, then, rational self-interest generally represents the exclusive principle of action for AI. My decision to invade my neighbor's land and thereby inflict injury upon him is a calculation made solely with a view towards what I consider is best for me and not informed by anything like Sympathy or Justice. If I choose not to invade, it must only be because I expect either that he will invade me in retaliation or that I expect for some other reason to receive greater long-term benefit for taking a peaceful policy. In the primacy of rational self-interest, then, we also see the close applicability of the Hobbesian moral framework to computer intelligence.



*Equality and Individualism*

Finally, this leads us to consider the radical individualism and equality of physical and mental condition that characterizes humans in the State of Nature. The two premises are closely related and represent a dramatic reversal of the traditional understanding of human nature.

Hobbes sees man as fundamentally isolated from family and society as well as conflict-prone, not least because of the differing appellations of words which causes misunderstanding. Contrast this with Aristotle who held that "man is by nature a social animal" and so "society is something that precedes the individual."[89] Hobbes's Ninth Law of Nature prohibits Pride and obliges "every man to acknowledge the other for his Equal by Nature"[90] since, unless men were equal, they would never enter into mutual Peace, but would find that the self-interest of the strong would be to dominate their inferiors. At the same time, because "the difference between man, and man, is not so considerable, as that one man can thereupon claim to himself any benefit, to which another may not pretend, as well as he,"[91] there is a consequent mistrust or "diffidence" about the intentions of others. If I declare a certain plot of land to be my property, I will immediately suspect my neighbors, who are my equals in strength and ambition, of making designs upon it.

The connection with Hobbes's individualism is most strikingly observed in his view of the family within which "there is not always that difference of strength or prudence between the man and the woman, as that the right can be determined without War."[92] By positing the equivalence of the father and mother, he destroys the natural paternal authority associated with the father in the family. And so also with children, that authority "is not so derived from the

---

[89] Aristotle, *Politics*. 1253a.
[90] *Supra, Leviathan* at XV. 107.
[91] *Id* at XIV. 87.
[92] *Id* at XX. 139.



Generation, as if therefore the parent had Dominion over his child because he begat him, but from the child's Consent."[93] Rather than admitting the existence of some natural authority separate from the political, Hobbes claims that the dominion of a father over his children is of the same sort of that of the Sovereign over his citizens: a covenant entered into by consent in order to achieve self-preservation. This has the further effect of breaking the natural familial ties that would otherwise enable each member of the family to rely on every other member for support and protection. Equality thus begets individualism and both, in turn, make the State of Nature one of isolation, distrust, and paranoia.

That equality and individualism apply equally to AI, is very plausible. In some MAS, we can imagine agents that are completely equal in cognitive and active capacities. In a given experimental setup, every "instance" of a learning agent can be the result of the same code and thus share precisely the same decision processes, set of potential actions, and motivating principle. The only thing separating each agent would be the set of lived experiences unique to it and the accidents of chance and position that may cause individual deviances in observed behavior. And while individualism does not follow from equality of condition for AIs as it does for the Hobbesian man, it applies just the same, but for different reasons.

Intelligent agents have no concept of the family or natural authority that Hobbes goes to such lengths to deny for humans. Whereas one might reasonably consider that a father has a natural tendency to protect and provide for his child in the State of Nature, no such relation exists amongst agents in a MAS except by convention. For agents that behave in a rational, self-interested way, any tendency towards cooperation and mutual support that arises is calculated to achieve that ideal of individual self-preservation which Hobbes places at the foundation of every relation of dominion between father and child, father and mother, and citizen and Sovereign. Just as the premises of equality and independence produce distrust and violence in humans, so they may also potentially turn out to be a cause of defecting behavior in multi-agent systems. If this is the case, we should expect to see improvements in cooperation with the introduction of an agent with dramatically superior power to dominate or reward and thus shape the behavior of other agents (roughly a picture of natural authority).

---

[93] *Id* at XX. 139.

In summary, I have shown how, starting from Hobbes's materialist and determinist natural philosophy, he incrementally deduces the moral subjectivism, rational self-interest, radical individualism, and equality of condition that constitutes the human condition. Furthermore, I've correlated each premise with his infamous conclusion that the State of Nature is a state of "continual fear, and danger of violent death; and the life of man, solitary, poor, nasty, brutish, and short."[94] By drawing out the causal links in his deductive methodology, we clearly see that for subsequent philosophers to avoid his political conclusions, they must either reject his foundational assumption about a material and deterministic universe or else uncover a flaw in any of the four succeeding moral philosophical premises. Since these properties have been shown to be applicable to the standard form of AI methodology (with agents adopting a rational, self-interested viewpoint), we should reasonably expect to find Hobbes's description of a brutal State of Nature and his political prescription, the Sovereign, to be important components of the problem and solution for cooperation in MAS.

---

[94] *Supra*, *Leviathan* at XIV. 89.



**The State of War**

From the preceding model of human nature, Hobbes goes on to show that the state of pre-social man is a state of war where all involved lose out. In this, it is notable that he stands apart from other social contractarians like Locke and Rousseau who paint the State of Nature in almost idyllic tones preferable to their 18th century political societies. It is arguably the case that, for AI in social dilemmas like the Prisoner's Dilemma, the consequences of the cycle of defection make it far closer to the Hobbesian prediction. Consequently, there should be great interest in understanding his particular political solution, and before that, his diagnosis for why the State of Nature is a state of war and how it follows from human nature.

Hobbes begins by distancing himself from a long-standing strand of thinking about cooperation and human nature: "It is true that certain living creatures such as bees and ants, live sociably with one another (… by Aristotle numbered amongst political creatures), … and therefore some man may perhaps desire to know why mankind may not do the same."[95] For animals, the "social contract" is written into the very chromosomes of their cells; rationality is clearly not necessary for the emergence of cooperation. Hobbes believed, however, that not only was it unnecessary, but rational self-interest was the very source of the problem since "the agreement of these creatures is natural; but that of men is by covenant only, which is artificial."[96]

Humans do not naturally work together and furthermore, "the difference between man, and man, is not so considerable, as that one man can thereupon claim to himself any benefit, to which another may not pretend, as well as he." The result of equality of condition is *diffidence*, or distrust, amongst men, since "if any two men desire the same thing, which nevertheless they cannot both enjoy, they become enemies; and in the way to their End, endeavor to destroy, or subdue one another."[97] Faced with this position of vulnerability with respect to his fellow men, each man inevitably pursues the rational policy of pre-emptive invasion. Owing to the high potential for conflict in the State of Nature, rational individuals recognize that striking first in their struggle for life provides them with the necessary advantage to survive. Consequently, the State of Nature can be called a state of war and "Force and Fraud are in war, the two Cardinal virtues."[98] Propositions 3 and 5 make these seeming vices morally justifiable since each man has the Right of Nature "to use his own power, as he will himself, for the preservation of his own Nature; that is to say, of his own Life."[99]

---

[95] *Supra*. *Leviathan* at Pt. II, Ch. 17.
[96] *Id*

Of course, when everyone has a Right of Nature to seize whatever it is in one's power to obtain, human wellbeing is fleeting and building civilization is impossible. "In such condition, there is no place for Industry; because the fruit thereof is uncertain"[100] So then, if others are willing to lay down this right to achieve security, then the rational policy is always to seek peace as well. "But if other men will not lay down their Right, as well as he; then there is no Reason for anyone, to divest himself of his: for that were to expose himself to Prey, rather than to dispose himself to Peace."[101] But after enough experience of the state of war, there may come a point where a majority of individuals decide that they would rather submit to any authority or limitation on their power than go on any longer in that state. When such a plurality of agreement has been reached, each man, by "the mutual transferring of Right… which men call Contract,"[102] enacts what we call the Social Contract.

---

[97] *Id* at 87.
[98] *Id* at 90.
[99] *Id* at 91.
[100] *Id* at 89.
[101] *Id* at 92.
[102] *Supra, Leviathan* at 94.



Now, once enacted, what is necessary is the performance of the obligations of that contract "without which Covenants are in vain and but empty words; and the right of all men to all things remaining, we are still in the condition of War."[103] Hobbes finally introduces the role of the Sovereign, explaining that "there must be some coercive Power, to compel men equally to the performance of their Covenants, by the terror of some punishment, greater than the benefit they expect by the breach of their Covenant."[104] Only a figure with absolute and unquestionable authority could possibly coerce unscrupulous, opportunistic men to perform their contracts with one another. By "conferring all their power and strength upon one man" in an act which "is more than Consent, or Concord; it is a real Unity of them all," political society is finally established. The Commonwealth culminates in "one Person, of whose Acts a great Multitude, by mutual Covenants one with another, have made themselves every one the Author, to the end he may use the strength and means of them all, as he shall think expedient, for their Peace and Common Defense."[105]

From his model of human nature, Hobbes has produced a political argument for the state of war based on several key underlying factors: the inherent distrust that each agent has for the other, the consensus of a majority to limit their own rights in order to escape this plight, and subsequently, the necessity of an authority vested with the power to reward and punish agents in a way that makes them amenable to living in society. These are the three core components that comprise the force of Hobbes's solution.

---

[103] *Id* at 100.
[104] *Id* at 101.
[105] *Id* at 121.



# The Civilization Game

The Civilization Game is a game that mimics the agrarian conditions of pre-social humanity and represents an experimental model of the experience of primitive man in the Hobbesian State of Nature. Players move to acquire territory, farm rewards from existing territory, or invade other player's territory to loot their rewards. Long-term cooperative behavior (i.e. farming rather than invading) leads to overall higher cumulative rewards for all players, while defecting behavior leads to higher short-run rewards from invading, but lower long-run rewards since being invaded in turn incurs a much larger penalty. The game is turn-based and agents navigate a 2-dimensional grid-world, moving in the four cardinal directions. These actions translate into two broad categories of action, farming one's own land or invading others', that represent cooperation and defection, respectively. As they traverse the grid world, they acquire "territory" by labelling each square they occupy with their unique player ID. In order to mimic the incentives that exist in the State of Nature, an "invasion bonus," a positive reward given to the agent who has invaded another agent's territory, and an "invasion penalty," a negative fine levied on the agent who has been invaded, are instituted. Choosing to farm, meanwhile, produces a reward for the agent proportional to the amount of territory it owns on its turn (i.e. one point for each square held).

This setup captures the fundamental dynamics that Hobbes posits in the State of Nature. Man is isolated and self-sufficient, and, left to his own devices, can secure a reasonable subsistence for himself by farming his property. Once he finds himself in the company of neighbors and subject to a scarcity of resources, however, a social dilemma rapidly develops: "it comes to pass, that where an Invader hath no more to fear, than another man's single power; if one plant, sow, build, or possess a convenient Seat, others may probably be expected to come



prepared with forces united, to dispossess and deprive him, not only of the fruit of his labor, but also of his life or liberty. And the Invader again is in the like danger of another."[106]

Greed emerges as the incentive to invade others' territory, both because of the immediate ability to steal their goods, but also because of the expansion of one's own territory leads to more farming rewards each turn. Fear, also, is the incentive not to be invaded by another, for the same two reasons; it is also the incentive that is completely essential for Hobbes since the prospect of invasion or death is far more painful than the profits of invasion are pleasant. To reflect this, the invasion penalty is greater in magnitude than the invasion bonus.

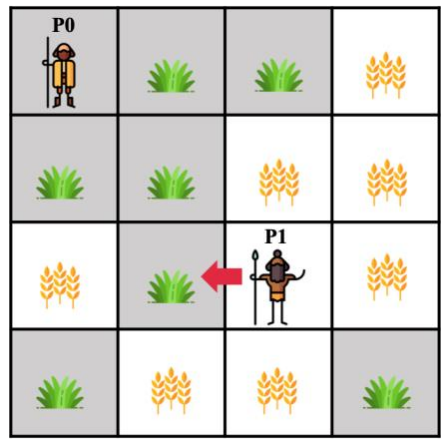

**Figure 5:** The Civilization Game with $b = 4$ and $p = 2$. Player $i = 0$ has name $P_0 = P0$ and Player $i = 1$ has name $P_1 = P1$. On his move $m = 1$, if $P1$ takes action $a = left$, then he has invaded $P0$'s territory. $P1$ will then gain a 25-point invasion bonus in addition to an 8-point farming reward and set $n_{t+1}^1 = True$ to indicate that $P0$ should incur a -10-invasion penalty on $m = 0$.

In practice, the grid-world is square sized (e.g. a 4x4 grid with 16 squares) and holds up to four agents, initially placed at the corners of the grid. Formally, for every moment in game time $t \in [0, 1, 2, \dots T]$, where we end the game after $T$ steps, the state is given by $\vec{s}_t \in S$, where $S$ is the set of all possible states, the action taken is $a_t \in A(\vec{s}_t)$, where $A(\vec{s}_t)$ gives the set of all legal actions available in state $\vec{s}_t$, and the subsequent reward is some real number $r_t \in \mathbb{R}$. For an instance of the Civilization Game with board size, $b$, and number of players, $p$, any state $\vec{s}_t$ can be represented as a single vector consisting of three components:

$$\vec{s}_t = \langle \vec{b}_t, \vec{n}_t, m_t \rangle$$


---

[106] *Id* at 87.

The vector $\vec{b}_t$ represents the $b \times b$ grid-world board, with each location containing either a player or a player's territory. $\vec{n}_t$ stores information about whether each player has been invaded in an array of booleans of size $p$ where the $i$th entry represents whether player $i$ has been invaded by another player since its last move. $m_t \in [0, 1, \dots, p-1]$ is an integer that tracks which player's move it is at time $t$ such that on move $m_t = m$, player $i = m$ takes an action. Player names that are placed in the locations of players in the board are stored in an array $\vec{P}$ of size $p$ (e.g. $[P0, P1, \dots P3]$ for a game with four players), and if a location on a board is the territory of player $i \in [0, 1, \dots p]$, then we simply place the integer $i$ at that location. So, in total, with a representation consisting of (1) a $b \times b$ board with each player appearing once, (2) a $p$ sized array tracking invasions, and (3) a single byte storing the current move, there are

$$\frac{b^2!}{(b^2 - p)!} \times bp(b-1) \times 2p \times p$$

possible legal states. For a game with a 3x3 board with 2 players, this translates into $|S| = 6{,}912$ possible states, but with a 4x4 board and 4 players, the state size balloons to $|S| = 67{,}092{,}480$ states.

The possible actions are any of the four cardinal directions, or if none are legally available, then the agent can choose to stay in its current location. Formally, $a \in [up, down, left, right, stay]$, and in a given time $t$ and state $\vec{s}$, we have $a_t \in A(\vec{s}_t)$ where $A(\vec{s}_t)$ returns the legal actions for player $i = \vec{s}_{t,m}$ that meet two criteria: (1) taking action $a$ will not



cause the agent to move out of the boundaries of the board, and (2) taking action $a$ will not cause the agent to move onto the location of another player. Because there are $p$ players in the Civilization Game, at each time $t$, the collective actions for each player are stored in an action array $\vec{a}_t = \langle a_t^0, a_t^1, \dots, a_t^{p-1} \rangle$. Importantly, the game is also asynchronous, or turn-based. So, while $\vec{a}_t$ contains an action for each player at every time $t$ for ease of notation, only the action for player $i = m_t$ actually executes at time $t$.

Traditionally, most standard MAS methodology utilizes a synchronous model where "agents repeatedly and simultaneously take action, which leads them from their previous state to a new one," but a problem in defining the transition functions of agents arises "due to the fact that the state in which the agent ends up after taking a particular action at a particular state depends also on actions and states of other agents." There are two standard ways to address this problem: we can "think of the transition function of the entire system, a mapping from the states and actions of all agents to the new states of all agents" or else we have to "define the transition function of each agent independently, and account for the effects of other agents by making the function nondeterministic."[107] Instead of taking either of these routes, agents play the Civilization Game in an asynchronous, turn-based way that allows the transition function to remain deterministic.

We define that transition function $T(\vec{s}_t, \vec{a}_t)$ as follows: (1) take the single action $a_t^i$ for player $i = s_{t,m}$ from location $l = INDEX(\vec{b}_t, P_i)$, (2) move the player from position $b_t^l$ to a new location $l' = MOVE(b_t^l, a_t^i, i)$ defined by:

$$MOVE(\vec{b}_t, a_t^i, i) = \begin{cases} l - b & a_t^i = up \\ l + b & a_t^i = down \\ l - 1 & a_t^i = left \\ l + 1 & a_t^i = right \\ l & a_t^i = stay \end{cases}$$

And setting $b_{t+1}^{l'} = P_i$ (3) if another player's territory was in the location moved to, or $b_{t+1}^{l'} \in [0, 1, \dots p] - [i]$, then set that player's invaded state to be True by setting $n_{t+1}^{i'} = True$ for player $i' = P_{b_{t+1}^{l'}}$, (4) set the previous location to be the player's territory $b_{t+1}^{l'} = i$ unless $a_t^i = stay$, (5) if player $i$ was invaded, or $n_t^i = True$, then reset it to $n_t^i = False$, and finally, (6) increment the move to the next player's turn $m_{t+1} = (m_t + 1) \% (p - 1)$. Notice that we assume that $a_t^i \in A(\vec{s}_t)$ is taken from the set of legal actions available to agent $i$ at time $t$ so that we preclude the possibility of moving to a location outside the board or onto the same location as another player.

The reward a player receives on its turn in the Civilization Game is determined by three factors: the amount of territory the agent controls and can farm, the invasion bonus accrued if it invades its neighbor's territory (i.e. gains +10 points), and the invasion penalty if the agent was invaded itself by another player (i.e. loses -25 points if $n_t^i = True$). Formally, for player $i$ we can write the reward function as:

$$R_i(\vec{s}_t, \vec{a}_t) = \begin{cases} TERR_i(s_{t,b}) + 10\ INVADE_i(s_{t,b}, a_t^i, i) - 25\ INVADED_i(s_{t,n}) & s_{t,m} = i \\ 0 & s_{t,m} \neq i \end{cases}$$

where the amount of territory is calculated by $TERR_i(\vec{b}_t) = COUNT(\vec{b}_t, i)$, whether the player has invaded another player's territory is calculated by:

$$INVADE_i(\vec{b}_t, a, i) = \begin{cases} 1 & b_t^{l'} \in [0, 1, \dots p] - [i] \\ 0 & o/w \end{cases}$$

And whether the player has been invaded is determined by



$$INVADED_i(\vec{n}_t) = \begin{cases} 1 & n_t^i = True \\ 0 & n_t^i = False \end{cases}$$

Note that, defined in this way, the Civilization Game satisfies the Markov property since the deterministic transition from any state-action pair $T\colon S \times A \to S'$ does not depend upon prior states. Furthermore, it satisfies the definition of a Markov game $M = (S, O, \vec{A}, T, r)$ with a perfectly transparent observation function $O_i(\vec{s}_t) = \vec{s}_t$ and a reward function $r\colon S \times \vec{A} \to \mathbb{R}$. Because of the incentives for both greed and fear, the Game also contains an embedded Matrix Game Social Dilemma, but this must be proved experimentally by observing the long-term payoffs achieved by actual agent gameplay.

## Hobbesian Q-Learning

As I have already described, Q-Learning is a model-free reinforcement learning algorithm whose goal is to learn a policy that determines which actions to take under what circumstances. Normally, Q-Learning converges to an optimal policy when playing a finite Markov Decision Process,[108] but in non-stationary environments like the Civilization Game where other actors can interfere with state transitions and rewards, it generally can't distinguish between the effects caused by its actions and those caused by external forces, and consequently, fails to converge. In these types of unstable environments, there is no general theory for the performance of Q-Learning. In general-sum games there can be multiple equilibria and thus initial conditions can alter the long-run payoffs that an agent and others involved can earn. Consequently, in some classes of games, many different kinds of dynamics can converge on equilibria with low payoffs even though high payoff equilibria exist.[109]

The Civilization game represents an example of a game where a policy of invading offers high immediate incentives but results in negative long-run reward while a policy of farming may produce low rewards but leads to a positive long-run score. In a 4x4 grid world and 4 Q-Learning agents with the following value iteration update parameters: the discount factor $\gamma = 0.99$, the soft update parameter $a = 0.5$, and the annealing discovery parameter $\varepsilon = 0.1$, we observe that agents don't converge to a high payoff cooperative strategy (i.e. avoiding invasion in favor of farming, which leads to positive long-term rewards). Rather the learning agents hardly out-perform agents that behave purely randomly, thus invading one another regularly, and any improvements in score are sporadic and temporary.

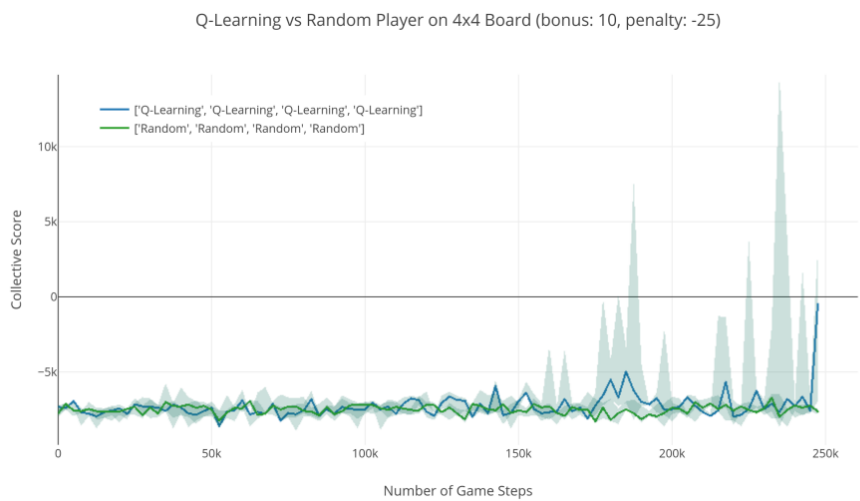

**Figure 6:** Collective scores of 4 Q-Learning agents playing on a 4x4 board versus collective scores of 4 Random agents. Collective score sums all players' rewards earned in a $bin = 2.5k$ game step period and divides by $bin$ to get a time average. Error bars are the max and min values from $trials = 3$ runs while the line represents the median scores.

From the Hobbesian game-theoretic perspective, Moehler succinctly summarizes this problem in the following way:



> If a situation of social interaction among rational individuals, who behave as if they were to
> maximize their expected utility, has the form of a one-shot PD game, cooperation does not take
> place, assuming common knowledge of rationality and complete information. Rational individuals
> alone are not able to realize the possible gains of cooperation in a one-shot PD game. Instead, they
> end up with a suboptimal outcome, both individually and collectively. To realize the (collectively)
> optimal outcome, an external authority, such as the state, is needed to transform the one-shot PD
> game into another game that makes mutual cooperation rational.[110]

In order to remedy this fundamental issue, we need to introduce additional social mechanisms that equip Q-Learning players with means of learning and interacting with other players in the game. Making learning agents social by implementing Hobbesian mechanisms for producing cooperation in MAS avoids explicitly requiring either a single programmer to coordinate and control all agents or, in the other extreme, allowing agents to behave in an unconstrained manner and appealing to a supervisor for resolving conflicts.[111] We call a learning agent with the basic Q-learning framework, but equipped with the Hobbesian mechanisms of 1) Opponent Learning Awareness, 2) Majoritarian Voting, and 3) Sovereign Reward-shaping, a Hobbesian Q-learner or HQ-learner.

*Opponent Learning Awareness*

The three following social mechanisms introduced closely follow the Hobbesian account of the formation of political society from the state of consistent defection that learning agents find themselves stuck in. The low payoff equilibrium of defection is, as Hobbes would call it, a state of war that "consists not in actual fighting, but in the known disposition thereto,"[112] transitions in his view to a situation in which each player is "willing, when others are so too, as

---

far-forth, as for Peace, and defense of himself he shall think it necessary, to lay down this right to all things, and be contented with so much liberty against other men, as he would allow other men against himself."[113] The problem facing an agent who would like to stop invading and being invaded, leading to negative rewards, is that as soon as he "lays down his arms," another agent has an incentive to invade him for short-term gain. "Rational individuals recognize, according to Hobbes, that striking first in their struggle for life provides them with the necessary advantage to survive in the State of Nature."[114] The Hobbesian term for this phenomenon is Diffidence, or the distrust of other player's intentions, and plays a central role in the unattainability of cooperation for him. This is also why Moehler calls the Hobbesian State of Nature an Assurance game and considers the primary role of the Sovereign to be one of eliminating Diffidence and institutionalizing social trust. "In order for rational individuals to leave the State of Nature, and in this sense to cooperate with each other, the Sovereign does not have to change the structure of the game that rational individuals play in Hobbes' State of Nature... Instead, the Sovereign only must assure the individuals that if they lay down their rights to everything and agree to keep their contracts, all other individuals will do so, too. *The Sovereign must only assure the individuals' trust in one another to establish society*."[115]

The first thing the Sovereign does is create a channel of communication that makes each agent's intentions accessible to every other player in the game. This Opponent Learning Awareness mechanism mitigates the mistrust and uncertainty surrounding intentions by shaping the anticipated learning of other agents in the environment. It modifies the learning rule to include an additional term $\Delta_t^i$ that accounts for the impact of agent $i$'s learning update on the

---

anticipated parameter update of the other agents. That term is the update from the Bellman equation:

$$Q_i(s_t, a_t) \leftarrow (1 - \alpha)Q_i(s_t, a_t) + \Delta_t^i$$

$$\Delta_t^i = \alpha(r_t + \gamma \cdot \max_a Q_i(s_{t+1}, a_t))$$

Every time $Q_i(s_t, a_t)$ gets updated on move $m = i$, we also update every other player's Q-table with $\Delta_t^i$. If, for example, a player $i$ chooses to take the defer action (i.e. cooperate and enter political society) in state $s_t$ and earns some reward $r_t$, which might be positive or negative, then the Sovereign enables every other agent to learn by applying the same update term $\Delta_t^i$ in $s_t$ but putting them "in the shoes" of player $i$ by adjusting the current move $s_{t,m}$ and the invasion store $s_{t,i}$ to reflect an alternate reality in which the other agent is the one deciding to take an action in the state.

In other words, OLA induces a sort of "empathy" in which, on every player's turn, all other players can "imagine" what it might be like to be in that situation and take that action, and their proclivity towards or away from the action is adjusted in the same direction as the player who actually does take the action. This mechanism is an internal feature of learning agents themselves, not an adjustment in the framework of the game or social dilemma. Formally, it modifies each player's Q-value update to adjust at every time step $t$ and every move $m$, rather than just on its turn $s_{t,m} = m = i$ for player $i$, and gives us the new update equation:

$$Q_i(s_t^i, a_t) \leftarrow (1 - \alpha)Q_i(s_t^i, a_t) + \Delta_t^m \text{ for } i \in [0, 1, \dots p - 1]$$

Where $\Delta_t^m$ is the learning update for the player whose move it currently is and

$$s_t^i = \langle SWAP(\vec{b}_t, INDEX(\vec{b}_t, P_i), INDEX(\vec{b}_t, P_m)), SWAP(\vec{n}_t, i, m_t), m_t = i \rangle$$

$$SWAP(\vec{v}, i, i') = \{\vec{v} | \ temp = v_i, v_i = v_{i'}, v_{i'} = temp\}$$



is the "alternate reality" state where player $i$ is put in the place of player $m$ by swapping board locations, invasion states, and current move.

As another consequence of this new OLA update, every player performs a Q-update every turn, rather than as before, each player performing its update only every $p$ turns. With this Hobbesian learning awareness mechanism built in, we should expect to see faster and steadier convergence to equilibrium policies across all players by increasing the frequency of learning and sharing those updates. It can still be the case that the equilibrium converged to may be a low payoff one, (i.e. mutual defection) so in order to reach the high payoff equilibrium, we need a further mechanism to ensure that the policy convergence for agents goes toward cooperation.

*Majoritarian Voting*

The only way to escape this State of Nature for Hobbes, is to form a political union under a Sovereign with sufficient instruments and power to shape the policy of individual agents. To erect this Commonwealth is "to confer all their power and strength upon one Man;" an act which "is more than Consent, or Concord; it is a real Unity of them all." This action is "the Generation of that great Leviathan, or rather of that Mortal God, to which we owe under the Immortal God, our peace and defense."[116] Practically, how does this political society come about? In the Hobbesian account, after a while, agents come to recognize the frustrating predicament in which they are stuck and, as we see in the Civilization game, rewards are consistently negative. What is required is to offer players a vote in order for them to signal to others their desire, whether to remain in or to escape the State of Nature. After the final tally, "if the representative consist of many men, the voice of the greater number, must be considered as the voice of them all."[117] Each agent votes to "authorize and give up my Right of Governing myself to this [Sovereign], on this condition, that thou give up thy Right to him and authorize all his Actions in like manner."[118] If a majority of agents vote to "defer" their right to choose their own action in favor of the Sovereign's choice, then for Hobbes, that vote is binding on all players since the threat of preemptive invasion disappears. "In a civil state, where there is a Power set up to constrain those that would otherwise violate their faith, that fear is no more reasonable, and for that cause, he which by the Covenant is to perform first, is obliged so to do."[119] With the Sovereign in power through that vote, invasion becomes *sine jure* because the Right of Nature no longer exists, and consequently the Sovereign enforces the social law that no player may enter another player's territory on its turn, but rather should stay on its own property and farm it, which leads to positive long-term payoff equilibrium. As Hobbes puts it, "the Sovereign assigns to every man a portion according as he, and not according as any Subject or number of them, shall judge agreeable to Equity and the Common Good."[120]

---

[116] *Supra*, *Leviathan* at XVII. 120.
[117] *Id* at XVI. 114.

The Majoritarian Vote mechanism lowers the threshold of the number of agents disposed to cooperative action to achieve system-wide cooperation and once reached, universalizes that policy by making it binding on non-cooperative agents. Instead of allowing a single invading player to upset the cooperative equilibrium, a simple majority vote can commit all other players to take the "defer" action, and once players become habituated in this way to choosing to cooperate, they learn from the reward feedback that the policy is good for them in the long run. Formally, we adjust the Civilization Game dynamics to account for a new potential action "defer" that, with a simple majority vote every $p$ turns, commits all players to taking the "defer"

---

action. The new action set is $a_t \in [up, down, left, right, stay, defer]$ and we increment the

number of possible moves to $m \in [0, 1, \dots p]$, making move $m = p$ the period where all $p$ agents

take the Majoritarian Vote for whether to institute the Sovereign. The Hobbesian transition

function becomes piecewise based upon the current move:

$$T^H(\vec{s}_t, \vec{a}_t) = \begin{cases} T^{VOTE}(\vec{s}_t, \vec{a}_t) & s_{t,m} = p \\ T(\vec{s}_t, \vec{a}_t) & s_{t,m} \neq p \end{cases}$$

Where $T(\vec{s}_t, \vec{a}_t)$ is the standard transition function from the standard Q-Learning model and the

voting procedure is described by:

$$T^{VOTE}(\vec{s}_t, \vec{a}_t) = \begin{cases} \vec{s}_t & VOTE(\vec{a}_t) > p/2 \\ \langle s_{t,b}, s_{t,n}, s_{t,m} = 0 \rangle & VOTE(\vec{a}_t) \leq p/2 \end{cases}$$

The function to count votes is given by $VOTE(\vec{a}_t) = \sum_{i=0}^{p-1}[a_t^i = defer]$. A successful vote

simply forces each player to take the "defer" action by setting $A(s_t) = [defer]$ until the next

vote at $t' = t + p$, not moving from their current board location since $T(s_t, defer) = s_t$, and

thus changing nothing about the state except taking $p$ turns. A failed vote simply advances the

move to player $i = 0$ but with $defer \notin A(s_t)$ for all players until the next vote at $t'$. The reward

each player accrues by playing "defer" after a successful vote is simply the farming bonus

proportional to the amount of territory held $R_i(\vec{s}_t, defer) = TERR_i(s_{t,b})$. As long as a majority

of agents choose to defer to the Sovereign, we should see an improved convergence to

cooperation in the MAS playing the Civilization Game.

*Sovereign Reward-shaping*

      With OLA, we have an intra-agent mechanism for modifying the way that agents learn

and share their intentions leading to faster and more stable convergence while with Majoritarian

Voting, we've introduced a modification in the game environment that pushes players towards

the high payoff equilibrium. In the Hobbesian narrative, we're halfway to the state of society:



individuals who are empowered to overcome their mutual Diffidence subsequently elect, by common assent, a single authority to whom they defer their actions. What remains is to institutionalize the behavior of this Sovereign and determine, from the Hobbesian perspective, how it conduces to the security and benefit of political society. Famously, Hobbes believed that the Sovereign exercised absolute power over its subjects and no policy it executed could be considered unjust. Even so, he still held that a good Sovereign knew that its own majesty and dominion went hand-in-hand with the flourishing of the state, for "the Nutrition of a Commonwealth consists in the Plenty and Distribution of Materials conducing to Life."[121] To this end, this primary prerogative is committed to the Sovereign: "the Power of Rewarding with riches, or honor; and of Punishing with corporal, or pecuniary punishment, or with ignominy every Subject according… as he shall judge most to conduce to the encouraging of men to serve the Commonwealth, or deterring of them from doing disservice to the same."[122] The institution of reward for players when they choose to defer to the Sovereign and penalty when they defect is the last component in the Hobbesian political prescription and requires modifying the Civilization Game.

In particular, after every $p$ turns, if the Majoritarian Vote succeeds, or $VOTE(\vec{a}_t) > p/2$, then every player receives a 10 point bonus and performs a Q-update for taking the $defer$ action in that state. Even players who didn't vote $defer$ still receive the bonus and perform the update *as if* they had voted to cooperate. If the Vote fails, then only those players who did vote to $defer$ receive a -10-point penalty in line with the Hobbesian and Assurance game view that those who attempt to lay down their arms while others choose to invade, suffer terrible consequences. Stated formally, we add a new integer component to the game state, $r \in [-1, 1]$, that represents

whether we should apply the Sovereign reward or penalty to players after a vote. The new game state is then:

$$\vec{s}_t = \langle \vec{b}_t, \vec{n}_t, m_t, r_t \rangle$$

And the transition vote function $T^{VOTE}$ only is modified to reflect the new state, since it is the only function that touches the new state component, not the overall Hobbesian transition $T^H$ or the standard transition $T$:

$$T^{VOTE}(\vec{s}_t, \vec{a}_t) = \begin{cases} \langle s_{t,b}, s_{t,n}, s_{t,m}, s_{t,r} = 1 \rangle & VOTE(\vec{a}_t) > p/2 \\ \langle s_{t,b}, s_{t,n}, s_{t,m} = 0, s_{t,m} = -1 \rangle & VOTE(\vec{a}_t) \leq p/2 \end{cases}$$

The reward function is also adjusted to account for the Sovereign's reward/punishment $R_i^S(\vec{s}_t, \vec{a}_t)$ function:

$$R_i(\vec{s}_t, \vec{a}_t) = \begin{cases} TERR_i(s_{t,b}) + 10\ INVADE_i(s_{t,b}, a_t^i, i) - 25\ INVADED_i(s_{t,n}) & s_m = i \\ R_i^S(\vec{s}_t, \vec{a}_t) & s_m = p \\ 0 & o/w \end{cases}$$

Where the Sovereign reward function is given by:

$$R_i^S(\vec{s}_t, \vec{a}_t) = \begin{cases} 10 s_{t,r} & a_t^i = defer \\ \max(10 s_{t,r}, 0) & a_t^i \neq defer \end{cases}$$

Finally, we utilize the same Q-update equation from OLA, but add the additional voting move $m = p$ to $i \in [0, 1, \dots p]$ with the further Sovereign reward from $R_p^S(\vec{s}_t, \vec{a}_t)$.

These three mechanisms, then, Opponent Learning Awareness, Majoritarian Voting, and Sovereign Reward-shaping form the backbone for a Hobbesian strategy for overcoming defection and entering and preserving the state of society. For Hobbes, each one is closely related to the others in this narrative and without all of them, we theoretically lose either the stability, speed, or direction of joint policy convergence for the MAS. Consequently, the HQ-Learning agent proposed will be deployed experimentally with a single, comprehensive toolkit.



## III.    Results

### Q-Learning versus Hobbesian Q-Learning

Having seen the problems that independent multi-agent reinforcement learning faces in overcoming the low payoff equilibrium to achieve the high payoff long-term cooperative policy and shown their similarity to the ones faced in the Hobbesian State of Nature, I have proposed that the three aforementioned political mechanisms offered by Hobbes should also be applicable solutions in MAS. The problem of mistrust and the ever-present fear of being invaded is mitigated by making each agent aware of its opponents' learning and using that to align all agents' policies. The next step is to lower the barrier for system-wide joint cooperation by means of making a majority vote for cooperative action binding on all agents in the game, thus also teaching the disruptive players about the benefits of cooperation. Finally, reward-shaping by the Sovereign ensures that when votes succeed, a central figure has authority to reward cooperation and punish defection.

In order to compare the performance of simple Q-Learning and Hobbesian Q-Learning, which is Q-Learning with the aforementioned Hobbesian mechanisms for shaping game and learning dynamics, the Civilization Game is played with $b = 4$ and $p = 4$ with an invasion bonus of 10 points and invasion penalty of -25 points as the experimental standard. The value iteration update parameters used are the discount factor $\gamma = 0.99$, the soft update parameter $a = 0.5$, and the annealing discovery parameter $\varepsilon_t = 0.9 \times 0.9999^t$. Majoritarian Voting activates the $defer$ action with a strict majority $VOTE(\vec{a}_t) > p/2$ and the Sovereign rewards $defer$ with a 15 point incentive while players who get duped (i.e. vote for $defer$ while other players decide to defect) incur a -10 point hit.



*Learning Curves*

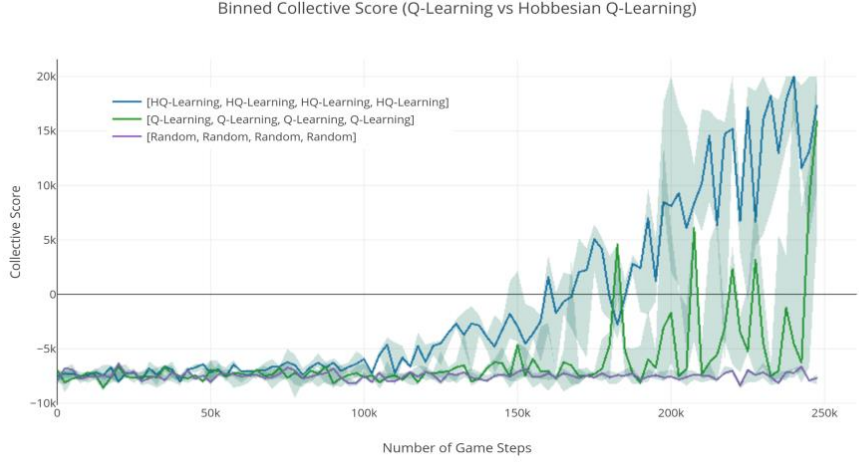

**Figure 7:** Collective rewards for all $p = 4$ players every $bin = 2500$ steps for $T = 250k$ steps.
Error bars indicate the results of running the game $trials = 3$ times with the median line
displayed. Score is a function of $TERR_i(s_{t,b})$, $INVADE_i(s_{t,b}, a_t^i, i)$, and $INVADED_i(s_{t,n})$.

The first point of comparison is the collective score earned over time. With a sample rate
of 0.01 and $T = 250k$ game steps, we get a bin size of $bin = 2500$. Collective Score earned
starting at game step $t$ is then defined as

$$CS_{t,bin}^B = \sum_{i=t}^{t+bin} \sum_{a=0}^{p} r_i^a$$

where $r_i^a$ is the reward earned by agent $a$ at time $i$. Random play consistently incurs negative
reward and acts as the base comparison. The rewards available to Q-Learning and Hobbesian Q-
Learning are made equal even though the learning remains distinct in order to make a fair apples-
to-apples comparison. Specifically, the rewards that the Sovereign provides apply to Q-Learners
if a majority decide to take the *defer* action, but unlike HQ-Learning, that reward does not
cause a Q-update. As we can see from Figure 7, the scores earned by Q-Learning players every
*bin* steps do improve over time, even jumping into positive territory a handful of times, but



learning is slow and sporadic as expected. For HQ-Learning, on the other hand, we see consistent and fast convergence to the high payoff equilibrium. The reward growth demonstrated in the learning curve is a dramatic improvement over independent reinforcement learning, but we still want to examine the policy underlying this performance improvement.

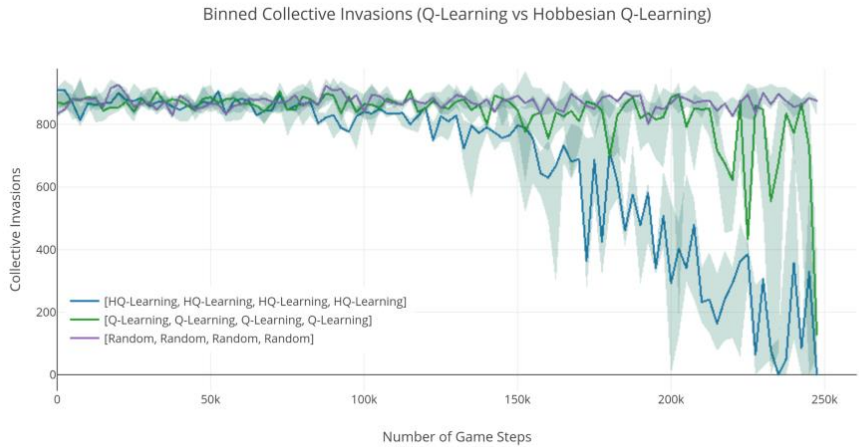

**Figure 8:** Collective invasions for all $p = 4$ players every $bin = 2500$ steps for $T = 250k$ steps. Error bars indicate the results of running the game $trials = 3$ times with the median line displayed. Invasions measured by tracking changes to the game state $\vec{n}_t$.

For any compelling case of cooperation in the Civilization Game, we should see that agents learn to stop invading one another. With the same game parameters and bin size of $bin = 2500$, we see that collective invasions drop off dramatically for HQ-Learning, converging to $0$ invasions per $bin$ steps. Collective Invasions starting at game step $t$ is measured every $p$ turns at times $E_t \in [t, t + p, \dots t + bin]$ as

$$CI_{t,bin}^B = \sum_{i \in E_t} \sum_{a=0}^{p} [n_i^a = True]$$

where $n_i^a = True$ if agent $a$ has been invaded at time $i$. As expected, invasions roughly inversely follow game score because of the invasion penalty that detracts from the score.



Even so, if agents are invading one another less often in HQ-Learning than standard Q-Learning, they must be taking some other action instead.

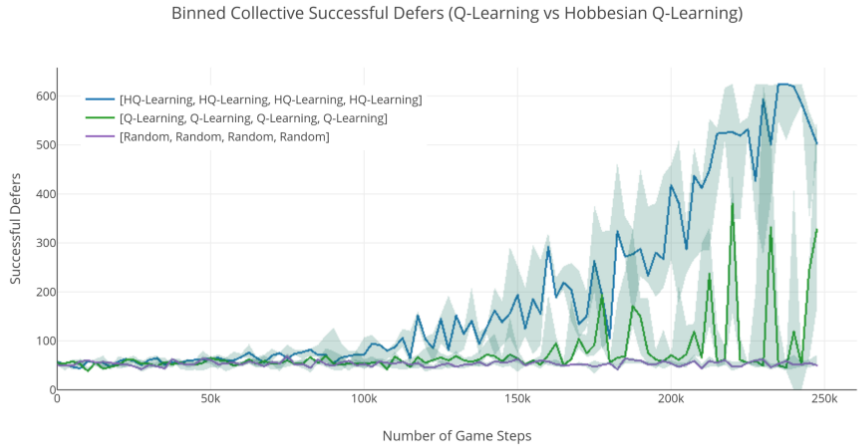

**Figure 9:** Collective successful defer votes for all $p = 4$ players every $bin = 2500$ steps for $T = 250k$ steps. Error bars indicate the results of running the game $trials = 3$ times with the median line displayed. Successful votes measured by tracking the game state when $r_t = 1$.

The other metric that shows that agents have learned to cooperate in the Civilization Game is the number of successful votes where more than half of agents choose to defer their actions to the Sovereign. Formally, collective successful defers is measured every $p$ turns at time $E_t \in [t, t + p, \dots t + bin]$ by:

$$SD_{t,bin}^B = \sum_{i \in E_t} [s_{t,r} = 1]$$

Just as in the collective score and invasions over time, Q-Learning learns to vote $defer$ only inconsistently while HQ-Learning learns the cooperative policy stably, because of the alignment of agent learning with OLA, and quickly, because of the Sovereign reward-shaping. OLA is especially important here because only cooperative action by $p/2$ agents at any time $t \in E_t$ is needed to induce a learning step for defecting agents that increases the Q-value of deferring at state $\vec{s}_t$. With more and more successful votes, defectors will



eventually discover that cooperating is more beneficial, even without necessarily having

taken the $defer$ action.

*Policies*

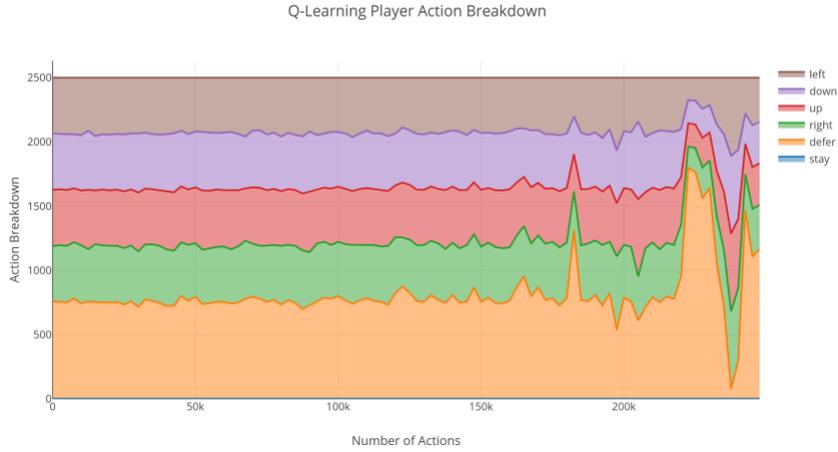

**Figure 10:** Breakdown of actions played by Q-Learning player $i = 0$ every $bin = 2500$ game
steps. The policy choice apparently fails to converge, with significant variance in the number of
$defer$ actions taken. Note that the $stay$ action is only legal when the other four cardinal
directions are not.

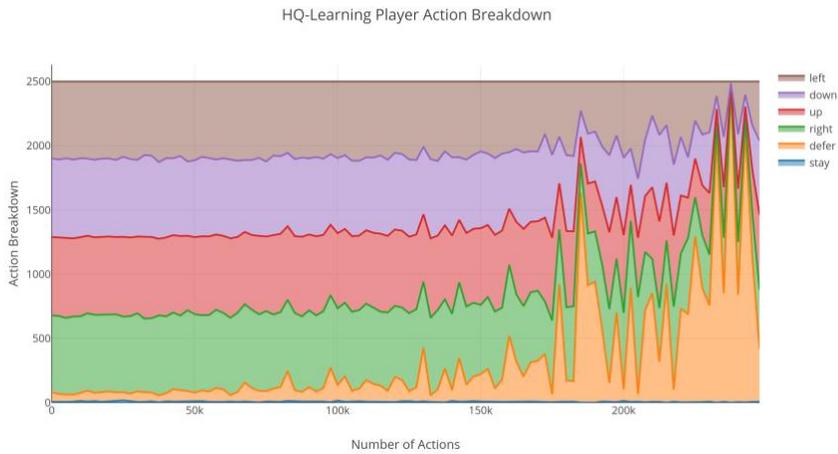

**Figure 11:** Breakdown of actions played by HQ-Learning player $i = 0$ every $bin = 2500$ game
steps. This player gradually learns the value of rejecting short-term incentives to navigate and
invade in favor of deferring to the Sovereign.



Lastly, to verify that the policy agents are learning is in fact to choose the Sovereign and political society rather than pursuing their own independent actions, we drop below the collective metrics and analyze the actions of a single agent playing the Civilization Game, arbitrarily chosen from the 4 players. Q-Learning vacillates between the high and low payoff equilibria as we anneal $\varepsilon$ over time while HQ-Learning learns to achieve the long-term cooperation policy by reliably deferring to the Sovereign.

**Matrix Game Analysis**

We can also perform an analysis to determine what kind of matrix game the Civilization Game SSD would translate into by playing agents with cooperative policies against agents who have learned to defect. As is the case for matrix games, Markov games like the Civilization Game become social dilemmas when the payoff outcomes $R, P, S$ and $T$ defined below satisfy the social dilemma inequalities. These outcomes are derived from:

$$R(s) := V_1^{\pi^C, \pi^C}(s) = V_2^{\pi^C, \pi^C}(s)$$

$$P(s) := V_1^{\pi^D, \pi^D}(s) = V_2^{\pi^D, \pi^D}(s)$$

$$S(s) := V_1^{\pi^C, \pi^D}(s) = V_2^{\pi^D, \pi^C}(s)$$

$$T(s) := V_1^{\pi^D, \pi^C}(s) = V_2^{\pi^C, \pi^D}(s)$$

Where $\pi^C$ and $\pi^D$ are cooperative and defecting policies respectively. The long-term payoff $V$ for player $i$ when the joint policy $\vec{\pi} = (\pi_1, \pi_2)$ is followed from an initial state $s_0$ is:

$$V_i^{\vec{\pi}}(s_0) = \mathrm{E}_{\vec{a}_t \sim \vec{\pi}(O(s_t)), \, s_{t+1} \sim T(s_t, \vec{a}_t)} \left[ \sum_{t=0}^{\infty} \gamma^t \, r_i(s_t, \vec{a}_t) \right]$$

with the temporal discount factor $\gamma \in [0,1]$. The social dilemma inequalities as previously defined are $R > P$ (mutual cooperation is preferred to mutual defection), $R > S$ (mutual



cooperation is preferred to being exploited by a defector), $2R > T + S$ (mutual cooperation is preferred to an equal probability of unilateral cooperation and defection), and either greed $T > R$ or fear $P > S$ (mutual defection is preferred over being exploited).

As discussed in the SSD section, we can define what a cooperation versus defection policy is by thresholding a continuous social behavior metric $\alpha : \Pi \rightarrow \mathbb{R}$ with values $\alpha_c$ and $\alpha_d$ such that $\alpha(\pi) < \alpha_c \leftrightarrow \pi \in \Pi^C$ and $\alpha(\pi) > \alpha_d \leftrightarrow \pi \in \Pi^D$. An obvious candidate for this metric would be invasions per hundred moves; if we set the threshold to $\alpha_c = 5$ and $\alpha_d = 15$, then plugging in the Q-Learning and HQ-Learning policies learned after $T = 250k$ steps, $\pi^Q$ and $\pi^{HQ}$, we have $\pi^Q \in \Pi^D$ and $\pi^{HQ} \in \Pi^C$. Having shown that the standard Q-Learning policy is a defecting policy and the HQ-Learning policy is cooperative, we can now play pairs of these policies against one another and calculate the long-term payoff $V$ and thus the matrix payoff values. For example, we can find the mutual cooperation payoff $R = V_0^{\pi^C, \pi^C}(s_0)$ by playing two HQ-Learning agents against one another. With $T = 100k$ game steps and the same game dynamics and payoffs as before, we get the payoff matrix:

| Civ Game | C | D |
|----------|---|---|
| C | R = 0.459, R = 0.459 | S = 0.426, T = 0.446 |
| D | T = 0.446, S = 0.426 | P = 0.455, P = 0.455 |

**Figure 12:** The Civilization Game SSD matrix game with long-term payoff $V_i^{\overline{\pi}}(s_0)$ divided by $T = 100k$. These values characterize the embedded MGSD within the Game.

In this run, by calculating $Fear = P - S = 0.029$ and $Greed = T - R = -0.013$, we see that the Civilization Game, and, to the degree that it conceptually resembles it, the Hobbesian State of Nature, is a Stag Hunt MGSD. However, the matrix payoffs change each trial run because of the existence of stochasticity in the discovery parameter, $\varepsilon$. The parameter is retained in order to prevent play from being "stuck" in a single state-action cycle, somewhat in the spirit of the



trembling hand perfect equilibrium[123] that also accounts for some randomness. Consequently, 15 trials are run to calculate different payoff matrices.

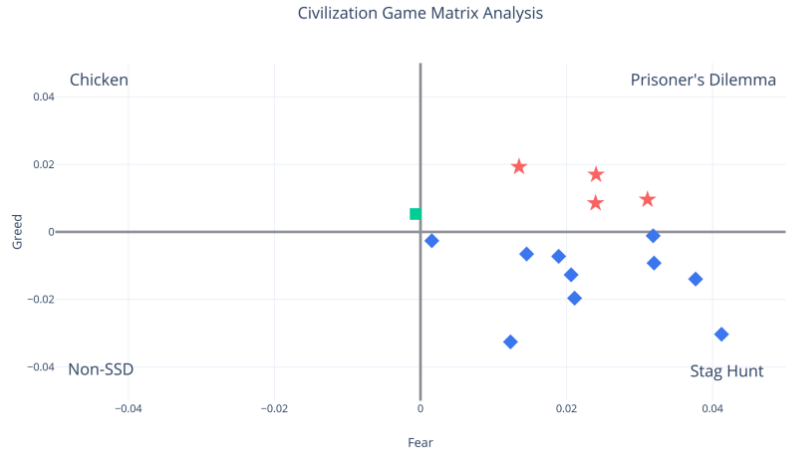

**Figure 13:** Civilization matrix game analysis with $trials = 15$ runs where each run consists of three match-ups: $HQL$ vs $HQL$, $HQL$ vs $QL$, and $QL$ vs $QL$ with $T = 100k$ steps. The resulting $V_i^{\vec{\pi}}(s)$ long-term payoff values are divided by $T$ for scale, then used to calculate $Fear$ and $Greed$.

The results for the greed and fear calculations differ by around 5% and two-thirds of runs end up as Stag Hunts. The fact that almost all runs have at least some fear factor while only one-third have a positive greed factor can be seen in the game reward structure: players are only rewarded 10 points for an invasion while being invaded incurs a -25-point penalty that is over twice as large. In addition, the existence of alternative channels for reward (i.e. farming and deferral bonuses) means that an invasion penalty only consists of about a third of the maximum possible reward on a given turn.

IV.    Conclusion

## Discussion

The central goal of this thesis has been to offer a new, interdisciplinary solution to the problems involved in producing cooperation in multi-agent reinforcement learning by looking to the way in which humans learn to cooperate in political society in the presence of incentives to defect. After implementing a new SSD and learning mechanisms inspired by that comparison, we see a surprisingly large improvement in game performance and convergence to a joint cooperative policy when compared to standard learning algorithms. Individually, there has been theoretical work in the field of MAS proposing frameworks similar to opponent learning awareness[124], learning sharing[125], reward shaping[126], and voting[127], but approaching the problem from a Hobbesian perspective merges and motivates them in a way that produces real experimental results. Specifically, the addition of an opponent learning awareness term $\Delta_t^i$ to each player's Q-update helps share information about how the behavioral intentions of other players changes and enables stable policy convergence. Secondly, the institution of a binding majoritarian voting procedure reduces the threshold of cooperative players needed to achieve a joint cooperative policy and thus increases the speed of convergence. Finally, when agents successfully vote to defer their actions, the Sovereign is empowered to disburse reward and punishment in a way that shapes the learning of defecting players.

The interdisciplinary application of solutions from Hobbesian political philosophy to the problem of cooperation in MAS is also unprecedented. While some philosophers interested in a computational theory of mind have seen connections to Hobbes's comment that "when a man

reasoneth, he does nothing else but conceive a sum total, from addition of parcels; or conceive a remainder, from subtraction of one sum from another,"[128] no attempt has been made to instantiate his insights in an experimental setting. In order to motivate that application, I've also offered a new argument that applies Hobbes's natural and moral philosophy, specifically his materialist, determinist, morally subjectivist, rationally self-interested, individualist, and egalitarian understanding of human nature, to artificially intelligent agents. This falls in line with, although certainly doesn't justify, Haugeland's dramatic assertion that Hobbes was "prophetically launching Artificial Intelligence"[129] when he published *Leviathan*. The argument raises questions about the degree to which AI agents and humans share the same nature and where differences might lie. Importantly, Hobbes also emphasizes the significance of the passions like vainglory, contempt, courage, and benevolence that don't fit neatly into existing computational frameworks.[130] This is certainly an exception to the otherwise striking comparison, but I will note that, for Hobbes, the passions are merely components of a psychology that is still materialist and completely determined by sense perception. Consequently, there is theoretically no reason why they couldn't be incorporated into a learning agent's computational decision process.

The Civilization Game SSD introduced in this thesis also offers a new blend of incentives and punishment meant to reflect the social dynamics that exist in the State of Nature and a unique opportunity to test Hobbes's philosophical framework. This experimental analysis contributes to contemporary Hobbes scholarship that understands it as a Stag Hunt, otherwise known as the Assurance Game. Moehler famously claimed that: "the situation of prudent individuals in Hobbes' State of Nature is best modeled by an assurance game, [because] the highest payoff for her is not achieved by unilateral defection, as suggested by the one-shot Prisoner's Dilemma game, but by mutual cooperation."[131] Skyrms also argued that if the shortsighted Foole were to consider the 'shadow of the future' in a repeated game, the alleged Prisoner's Dilemma game would be transformed into an assurance game.[132] Gauthier, also, believes that the adequate long-term representation of the game that rational individuals play in Hobbes' State of Nature is an assurance game, although he does not call the game by its name.[133] Fascinatingly, however, we can see from this thesis that it isn't always or necessarily an Assurance Game, but rather, depending on the way one adjusts the ratio between how tempting the invasion bonus is and how painful the invasion penalty is, the game might transform into a Prisoner's Dilemma instead.

---

[128] *Supra*, Leviathan at V.
[129] Haugeland, J. *Artificial intelligence: the very idea*. 1985. 23.
[130] *Supra*, Leviathan at VI.

Importantly, the matrix game analysis presented here is the first attempt, to my knowledge, to go beyond a close exegesis of Hobbes to an SSD analysis of the game-theoretic dynamics of the State of Nature. While scholars have studied Hobbes quantitatively via an iterated matrix game analysis[134], those investigations do not account for the fact that 1) the complexity of implementing cooperative versus defecting policies may be unequal, 2) successfully implementing a policy often requires solving coordination sub-problems, and 3) cooperativeness is a gradated quantity, not an atomic choice of action. Utilizing the Civilization Game SSD enables us to understand how agents in a realistic setting learn to choose and implement policies, and with the help of Hobbesian mechanisms, achieve a cooperative equilibrium.

## Future Work

This thesis illustrates the promising practical value of combining knowledge from the fields of computer science and political philosophy in understanding the evolution of cooperation and lays out some of the experimental groundwork to make that possible. Even so, a great deal remains for future work. Within just the bounds of the research question considered here, there would be great value in taking the time to implement a more state-of-the-art RL algorithm like deep Q-networks that utilize the same theoretical framework as Q-Learning but offers scalability to larger state spaces. This would open the door to testing more complicated games with more players and a bigger grid-world to explore. The dynamics of an 8 player Civilization Game, for example, may look very different from the 4 player and 2 player scenarios tested here.

There are also additional ideas in Hobbes that could be translated into promising mechanisms, like the detailed process for deriving and enforcing binding Covenants between different players and between players and the Sovereign. Implementing the Sovereign as a separate agent itself instead of a component of the game environment might also allow a researcher to analyze its interactions with other players and ask, for example, how good the Sovereign must be at punishing defection before the MAS devolves back to the low payoff equilibrium. Finally, varying the reward structures within the Civilization Game including the farming bonus, invasion bonus, and invasion penalty as well as the Sovereign reward-shaping incentives may likely result in different embedded MGSDs and consequently, different learning behavior.

Although here the three Hobbesian political mechanisms are considered together as a unified strategy, further MAS research may analyze the individual merits of each considered separately. Whether individually or taken as a whole, it would also be of great interest to observe



the performance of modified Hobbesian Q-Learning in other popular SSDs as well like the Wolf

Pack and Gathering games from Leibo[135], or especially the Stag Hunt-inspired Markov games

like Harvest and Escalation from Peysakhovich and Lerer[136]. We would expect to see similar

improvements in performance, especially for those games that preserve high level properties of

the Stag Hunt. The unique asynchronous turn-based dynamic of the Civilization Game might

also offer an opportunity to contrast experimentally considerations of player memory length and

complexity cost required to compute equilibria versus in the standard synchronous model.[137]

Generally, my hope is to encourage research in the field of AI or MAS to look beyond traditional

methodologies and search out interdisciplinary solutions, even in fields as unconventional as

political philosophy, to interdisciplinary problems.

Likewise, scholars of Hobbes, in line with his thoroughly empirical method, have at their

disposal a powerful new experimental tool that models the State of Nature in many important

respects, although admittedly simplified in others, and deploys Hobbesian, intelligent agents to

play that game. By varying the incentive structure in the Civilization Game, we can see how

varying certain underlying assumptions that Hobbes makes might affect his political conclusions.

I have already shown that increasing the invasion bonus / invasion penalty ratio transforms the

embedded MGSD in the Civilization Game from an Assurance game to a Prisoner's Dilemma.

Further research might also examine the consequences of reducing the effectiveness of the

Sovereign by either reducing the size of the reward or punishment or by making it apply only

once every $n$ times and observing whether agents still learn to cooperate. The results of such an

experiment might inform our perspective on Hobbes's infamous position that life under a

---

[135] *Supra*, Leibo et al.
[136] *Supra*, Peysakhovich et al.
[137] Bhaskar, V. *Asynchronous Choice and Markov Equilibria*. 1998. 1.



Sovereign, no matter how harsh, is always better than life in the State of Nature. This thesis also offers the first argument for the applicability of Hobbes's natural and moral philosophy to modern computational artificial intelligence. The obvious exception that has been highlighted is the role of the passions in directing human action that finds no direct corollary in mechanical RL. That being said, in Hobbes's mechanistic materialist psychology, there should almost certainly exist a way to mechanize the passions in his account in a way that makes them feasible to model or implement in computers.

In conclusion, it's clear that the evolution of cooperation is a fascinating and difficult problem of great practical importance for both humans and AI. Hobbes's own answer profoundly influenced the way in which humans organize our modern, liberal societies, and the solutions we develop to enable cooperation amongst artificially intelligent agents will similarly affect what technology in the 21st century will look like and how it will interface with us. There is good reason to be optimistic about a near future in which humans and computers work hand-in-hand to solve the biggest challenges we face, and my hope is that the work presented here might represent a small contribution to that larger project.